\documentclass[aps, prb, twocolumn, superscriptaddress, showpacs,
  notitlepage, reprint, citeautoscript]{revtex4-1}
\usepackage[utf8]{inputenc}
\usepackage{amsmath}
\usepackage{latexsym}
\usepackage{bm}
\usepackage{graphicx}
\usepackage{color}

\begin{document}

\title{Bilayer graphene under pressure: Electron-hole Symmetry Breaking, Valley Hall Effect, and Landau Levels}

\author{F. Munoz}

\affiliation{Departamento de F\'isica, Facultad de Ciencias,
  Universidad de Chile, Chile}
\affiliation{Centro para el Desarrollo de la Nanociencia y la
  Nanotecnología (CEDENNA), Santiago, Chile}

\author{H. P. Ojeda Collado} 

\affiliation{Centro At\'omico Bariloche and Instituto Balseiro,
  Comisi\'on Nacional de Energ\'ia At\'omica, 8400 Bariloche,
  Argentina}
\affiliation{Consejo Nacional de Investigaciones Cient\'ificas y
  T\'ecnicas (CONICET), Argentina}

\author{Gonzalo Usaj}

\affiliation{Centro At\'omico Bariloche and Instituto Balseiro,
  Comisi\'on Nacional de Energ\'ia At\'omica, 8400 Bariloche,
  Argentina} 
\affiliation{Consejo Nacional de Investigaciones Cient\'ificas y
  T\'ecnicas (CONICET), Argentina}

\author{Jorge O. Sofo}
\affiliation{Department of Physics and Materials Research Institute,
  The Pennsylvania State University, University Park, Pennsylvania
  16802, USA }

\author{C. A. Balseiro}
\affiliation{Centro At\'omico Bariloche and Instituto Balseiro,
  Comisi\'on Nacional de Energ\'ia At\'omica, 8400 Bariloche,
  Argentina}
\affiliation{Consejo Nacional de Investigaciones Cient\'ificas y
  T\'ecnicas (CONICET), Argentina}

\begin{abstract}
  The electronic structure of bilayer graphene under pressure develops
  very interesting features with an enhancement of the trigonal
  warping and a splitting of the parabolic touching bands at the K
  point of the reciprocal space into four Dirac cones, one at K and
  three along the T symmetry lines. As pressure is increased, these
  cones separate in reciprocal space and in energy, breaking the
  electron-hole symmetry. Due to their energy separation, their
  opposite Berry curvature can be observed in valley Hall effect
  experiments and in the structure of the Landau levels. Based on the
  electronic structure obtained by Density Functional Theory, we
  develop a low energy Hamiltonian that describes the effects of
  pressure on measurable quantities such as the Hall conductivity and
  the Landau levels of the system.
\end{abstract}

\maketitle
\section{Introduction}

Although the electronic structure of graphene has been known for more
than half a century, it was the pioneer work of Novoselov and Geim in
2004 what triggered an impressive scientific and technological
activity in this two-dimensional
system\cite{novoselov_electric_2004}. Indeed the exceptional
characteristics of graphene were fully revealed only after its
systematic isolation, characterization and the first studies of its
unusual mechanical, optical and transport
properties.\cite{Zhang2005,castro_neto_electronic_2009,Balandin2008,Bonaccorso2010,Novoselov2005,Katsnelson2006,Novoselov2007,Geim2007,DasSarma2011}
The band structure~\cite{Wallace1947} includes two points at the
corners K and K$^{\prime}$ of the Brillouin zone (BZ) that engender
the peculiar low energy properties of the material. In fact, the low
energy excitations around these points--the Dirac points--are
described by chiral quasiparticles behaving as massless Dirac fermions
leading to a number of remarkable
phenomena.\cite{castro_neto_electronic_2009,DasSarma2011,Beenakker2008a,Goerbig2011}

Bilayer graphene (BLG) in the Bernal stacking consists of two graphene
layers where only one of the two carbon atoms of the unit cell of the
top layer lies on top of an atom of the bottom layer. The bilayer unit
cell has four carbon atoms leading to four $\pi$-bands. Two of them
touch each other at the Dirac points, having a parabolic dispersion
relation with opposite curvature around it, which results in a zero
band gap semiconductor.  This simple-looking band structure wraps
surprising properties that make BLG an exciting material from the
point of view of exploring new physics and because of its potentials
for technological applications.

The unique features emerging from the BLG electronic structure are due
to different properties of the material.  On the one hand an electric
field perpendicular to the layers generates a tunable band gap
\cite{castro_biased_2007,McCann2006, Min2007,Taychatanapat2010}, a
required effect to engineer carbon-based semiconducting devices. On
the other hand, from the topological point of
view,~\cite{Hasan2010,Martin2008} the Bloch wavefunctions of BLG
present a rich behavior. Indeed by combining these two properties, the
detection of the predicted generation of pure valley
currents,~\cite{Rycerz2007} or valley-Hall
effect,~\cite{xiao_valley-contrasting_2007} has been recently
reported~\cite{Sui2015}. The topological properties are determined by
the Berry phase resulting from the winding of the phase of the Bloch
wavefunction along a close path around the Dirac
point.~\cite{Hasan2010,xiao_berry_2010,Ando2013,Ren2015a} The
consequences of a non-trivial Berry phase are diverse, in particular
the mentioned valley-Hall effect results from the Berry curvature.  In
the presence of an external magnetic field, the structure of the
Landau Levels (LLs) with a two-fold orbital degeneracy of the
topologically protected zero energy ($n=0$) levels of each valley is
also due to the structure of the Berry phase.

Trigonal warping effects on BLG have also been extensively studied
\cite{castro_neto_electronic_2009,tw_falko_altshuler_2007,tw_cserti_2007,tw_koshino_2009}. They
change the spectrum at low energies qualitatively generating four
Dirac cones with zero energy around the K and K$^{\prime}$ points of
the BZ. This is however a very small correction changing the spectrum
in an energy range of a few meV around the Dirac points. It has been
shown that distortions of the structure of BLG can be used to control
and enhance the splitting of the parabolic band-contact point into the
four cones that move away from the Dirac points
\cite{montambaux_merging_2009,de_gail_magnetic_2012}. The way this
splitting occurs is associated to topological invariants that define
the LLs spectrum in the presence of an external magnetic field
\cite{fuchs_topological_2010} unveiling the richness of the electronic
properties of BLG.

Here we present results for BLG under high pressure.  We show that
pressures in the range of $10$ to $100$ GPa modify in a substantial
way the band structure around the Dirac points with the corresponding
change of the wavefunctions and its topology. This is so because
pressure changes the interlayer distance increasing the coupling
between layers. In the resulting band structure, the two parabolic
bands evolve to generate four Dirac cones. One of them, with a marked
trigonal warping is centered at the K (K$^{\prime}$) point and three
elliptic cones move away towards the $\mathrm{\Gamma}$ point. The new
ingredient, which has been overlooked in the past, is the breaking of
the electron-hole symmetry.  The energy of the trigonal cone apex at K
is smaller than the energy of the apex of the three elliptical
ones. This difference can be of the order of $0.1$ eV and the Fermi
surface of neutral BLG under pressure consists of small electron and
hole pockets. In such a case, a large enough electric field opens an
indirect gap. In other words, pressure induces a Lifshitz transition
changing the nature of the Fermi surface that, depending on doping,
may include electron and hole pockets, each one having a
characteristic topological invariant (winding number).
Although the pressure needed to observe these effects is rather high,
this is still in the experimentally accessible range as, for small
samples, diamond anvil cells can reach
pressures\cite{mcmillan_new_2002} much higher than the ones used in
the present work.

We present the band structure and its evolution with pressure as
obtained with Density Functional Theory (DFT). In particular, we
analyze in detail the effect of pressure around the K and K$^{\prime}$
points of the BZ and present a tight-binding model that properly
describes this effect. The microscopic parameters are obtained by
fitting the DFT bands. We show that while the resulting electronic
structure is topologically trivial, in the presence of an electric
field the bands acquire a Berry curvature that lead to a non trivial
pressure dependent valley Hall effect. We also present results for the
LLs of BLG under pressure in the presence of an external magnetic
field. These effects can be measured, adding to the rich behavior
shown in BLG without pressure \cite{novoselov_unconventional_2006}.

\section{DFT calculations}

\begin{figure}[tb]
  \includegraphics[width=0.9\columnwidth]{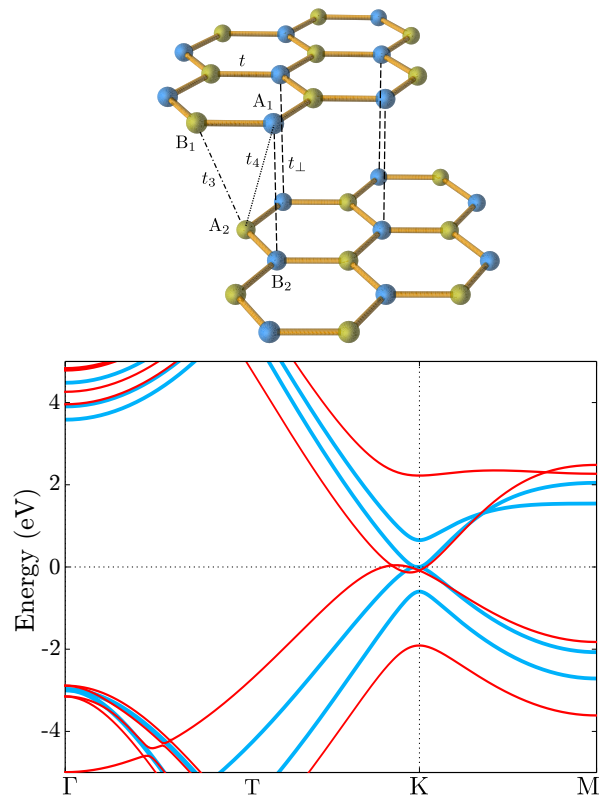}
  \caption{(Color online) Top: The atomic structure of bilayer
    graphene in the Bernal stacking. Carbon atoms that are on top of
    each other are depicted in blue, while those aligned with the
    hexagonal hole in the other plane are depicted in green. Also
    shown in the figure are the hopping parameters used in our tight
    binding model. Bottom: Band structure along the
    $\mathrm{\Gamma\rightarrow K\rightarrow M}$ path for two different
    pressures ($12$ and $96$ GPa, thick ligth blue and thin red
    curves, respectively). There are four bands around the Fermi level
    (horizontal dashed line); two bands whose separation at K
    increases with increasing pressure and two that overlap (see
    Fig. \ref{fig:BandsFull}).}
\label{bernal}
\end{figure}


We will assume the Bernal stacking for which the unit cell has four
carbon atoms, two in each plane. One of them is directly on top of an
atom in the other plane (labeled $A_{1}$ and $B_{2}$ in
Fig.~\ref{bernal}) while the other two are on top of the hexagonal
hole of the other plane (labeled $A_{2}$ and $B_{1}$ in the
figure). We use DFT to determine the change of the atomic positions
and the corresponding band structure with pressure, assuming that the
effect of pressure is to reduce the distance between planes. To test
the validity of this approximation, we used a more refined protocol
considering that the isotropic pressure exerts equal force on each
atom. This procedure is implemented by fixing the distance between
atoms $A_{1}$ and $B_{2}$ and changing symmetrically the position
perpendicular to the layers of atoms $A_{2}$ and $B_{1}$ (those
opposed to the hexagon center on the other layer) until the force on
them becomes equal to the force on the fist pair. This procedure
provides the structure corresponding to a uniform stress. We found
that the difference between the results of this more refined procedure
and those obtained using the rigid plane configuration was less than
$0.025$~\AA~at the largest pressure considered, leading to a
negligible effect on the band structure. In what follows we present
results from both procedures indistinguishably.  The conversion from
plane separation to pressure was determined from the forces on the
atoms.


The DFT calculations were done using a plane wave basis set as
implemented in the VASP code
\cite{kresse_ab_1993,kresse_ab_1994,kresse_efficiency_1996,kresse_efficient_1996}
with an energy cutoff of $450$ eV. The core electrons were treated
with a frozen Projector Augmented Waves (PAW)
scheme\cite{blochl_projector_1994,kresse_ultrasoft_1999}. The exchange
and correlation functional was approximated with a simplified
Generalized Gradient Approximation (GGA) as parametrized by Perdew,
Burke, and Ernzerhof
(PBE)\cite{perdew_generalized_1996,perdew_generalized_1997}. Spin
polarized calculations showed that in the range of experimentally
accessible pressures the system is non-magnetic. Spin-orbit coupling
was not includes in this first runs. Given that the distance between
the planes was a controlled parameter, not derived from total energy
minimization, the functional did not include van der Waals
corrections.

Additional calculations were carried out to determine if the Bernal
stacking is still the most stable BLG arrangement under pressure. We
found that the Bernal stacking always corresponds to the minimum
energy configuration---we will limit to this case from
hereon. However, it is worth mentioning that a configuration with a
lateral displacement with two top atoms symmetrically placed on top of
a hexagon is very close in energy ($\Delta E = 15$ meV/atom at $71$ GPa).

The important changes in the electronic structure of the bilayer under
pressure occur in a very small region around the K point, which
imposes a challenge to the numerical sampling of the Brillouin zone. A
sampling of $300\times300$, that in many cases would be enough to
guarantee convergence in terms of BZ integration, is not enough to
produced a reasonable determination of the Fermi level in the bilayer
under pressure. To overcome this issue we used a non-uniform sampling
with the equivalent of a $3000\times3000$ sampling in the neighborhood
of the K point and a $300\times300$ grid in the rest of the BZ.

\begin{figure}[ht]
  \includegraphics[width=0.9\columnwidth]{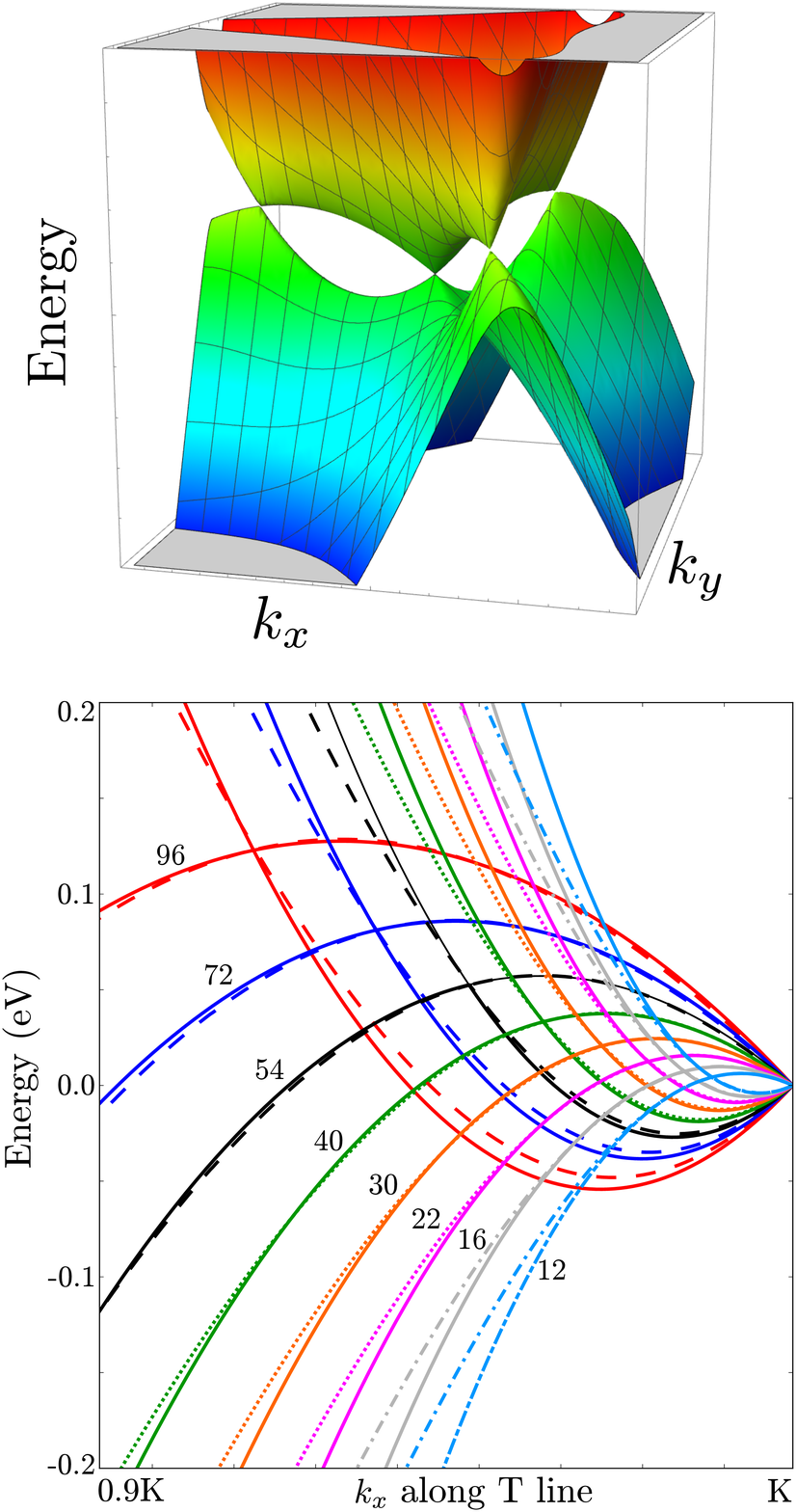}
\caption{Low energy band structure near the K point. Top: Band
  structure along the $\mathrm{\Gamma\rightarrow K}$ path for
  different pressures (indicated in the figure in GPa). The effect of
  the trigonal warping and the electron-hole asymmetry increase as
  pressure is increased. The dashed lines show the fitting with the
  tight binding model described in Sec.~\ref{sec:tight-binding}.
  Bottom: Detail of the four Dirac cones for a pressure of $96$
  GPa. Note that the energy of the Dirac point K (the central cone) is
  different from the other three.}
\label{fig:BandsFull}
\end{figure}

The band structure of BLG in the Bernal stacking is shown in the
bottom panel of Fig.~\ref{bernal} for the lowest and highest pressures
considered in this work. Around the K and K$^{\prime}$ points of the
BZ, there are four low energy bands. At zero pressure, two bands touch
with a parabolic dispersion relation at the Fermi level and two are
separated by an energy proportional to the inter-plane hopping matrix
element ($t_\perp$). As pressure is increased, the latter two bands
separate more and more from the Fermi level, while the other two
undergo a trigonal warping.

To display this trigonal warping in more detail, we plot the band
structure for increasing pressure in the top panel of
Fig.~\ref{fig:BandsFull}. With increasing pressure the parabolic
dispersion relation of the low energy bands becomes a set of four
Dirac cones,\footnote{It should be emphasize that trigonal warping
  effects are always present, even a zero pressure. However, in such a
  case they occur at much lower energy, since the $t_3$ and $t_4$
  hopping parameters (see next section) are rather small. In the first
  approximation, we take this limit as if $t_3=t_4=0$.}  one that
remains at K and three that shift in reciprocal space along the T path
of the BZ. The energy of the Dirac points along the T lines increases
with increasing pressure with respect to the point at K, reaching a
value close to 100~meV at the highest pressure considered. A
three-dimensional view of the bands at this pressure is shown in the
bottom panel of Fig.~\ref{fig:BandsFull}.

The trigonal warping was mentioned before in tight binding models of
the BLG
\cite{mccann_electrons_2007,nilsson_electronic_2008,castro_neto_electronic_2009}.
However, in all this models the Dirac cones along T remain at the same
energy as the cone at K. As it is evident from the DFT results, with
pressure there are extra ingredients that are missing in these tight
binding representations. These important parameters are discussed in
the next section where we present our tight binding model. The notable
effect to be included is the electron hole symmetry breaking that
enables the possibility to observe the Dirac cone splitting. Among its
consequences are the charge transfer between sublattices $A$ and $B$
as well as the unveiling of interesting and measurable topological
effects.

\section{Tight Binding Model}
\label{sec:tight-binding}
A tight-binding approximation for the four bands close to the Fermi
level of BLG has been discussed previously in the literature
\cite{mccann_electrons_2007,nilsson_electronic_2008}. We will use it
now to describe the effect of pressure.  As the latter increases, we
will see that the hopping integral $t_4$ between atoms $A_2$ and $A_1$
(see Fig.~\ref{bernal}) plays an increasingly important role: it is
responsible for the electron-hole symmetry breaking and consequent
energy separation of the Dirac cones at T with respect to the cone at
K. This important effect has been overlooked in the literature so far,
mainly because the effect is negligible at low pressure.

The four bands around the Fermi level, which derive from the $p_z$
orbitals, are described by the Hamiltonian
$\mathcal{H}=\mathcal{H}_{1}+\mathcal{H}_{2}+\mathcal{H}_{12}$ where
the first terms describes the electronic structure of the two graphene
layers and the last one includes the inter-plane coupling
\begin{eqnarray}
\nonumber
\mathcal{H}_{i}&=&-\sum_{\bm{k},\sigma}(-1)^{i}\left[(\varepsilon+V)\,a^{\dagger}_{i\bm{k}\sigma}
a^{}_{i\bm{k}\sigma}
+(-\varepsilon+V)\,b^{\dagger}_{i\bm{k}\sigma}b^{}_{i\bm{k}\sigma}\right]\\
&&-\sum_{\bm{k},\sigma}t\left[\phi({\bm{k}})\,a^{\dagger}_{i\bm{k}\sigma}b^{}_{i\bm{k}\sigma}+\phi({\bm{k}})^{*}\,b^{\dagger}_{i\bm{k}\sigma}a^{}_{i\bm{k}\sigma}\right]\,,
\end{eqnarray}
with $i=1, 2$. Here, $a^{}_{i\bm{k}\sigma}$ and $b^{}_{i\bm{k}\sigma}$
destroy electrons with wavector $\bm{k}$ and spin $\sigma$ in
sublattices $A$ and $B$ of the $i$-{th} plane, respectively,
$\varepsilon$ is the energy due to the charge transfer between the two
sublattices on each plane, and we have included an electric field
perpendicular to the BLG plane described by $V$. The matrix element
$t$ corresponds to the intra-plane hopping and
\begin{equation}
\phi(\bm{k}) = e^{iak_{y}} \left[1+2e^{-i\frac{3a}{2}k_{y}}\cos\left(
  \frac{a\sqrt{3}}{2}k_{x}\right) \right]\,,
\end{equation}
with the carbon-carbon distance $a=1.42$~\AA~.
The inter-plane coupling is described by
\begin{eqnarray}
\nonumber
\mathcal{H}_{12}&=&\sum_{\bm{k},\sigma}t_{\perp}\left(a^{\dagger}_{1\bm{k}\sigma}b^{}_{2\bm{k}\sigma}+b^{\dagger}_{2\bm{k}\sigma}a^{}_{1\bm{k}\sigma}\right)\\
\nonumber
&&+\sum_{\bm{k},\sigma}t_{3}\left(\phi({\bm{k}})\,b^{\dagger}_{1\bm{k}\sigma}a^{}_{2\bm{k}\sigma}+\phi({\bm{k}})^{*}\,a^{\dagger}_{2\bm{k}\sigma}b^{}_{1\bm{k}\sigma}\right)\\
\nonumber
&&+\sum_{\bm{k},\sigma}t_{4}\left(\phi({\bm{k}})^{*}\,a^{\dagger}_{1\bm{k}\sigma}a^{}_{2\bm{k}\sigma}+\phi({\bm{k}})\,a^{\dagger}_{2,\bm{k},\sigma}a_{1,\bm{k},\sigma}\right)\\
&&+\sum_{\bm{k},\sigma}t_{4}\left(\phi({\bm{k}})^{*}\,b^{\dagger}_{1\bm{k}\sigma}b^{}_{2\bm{k}\sigma}+\phi({\bm{k}})\,b^{\dagger}_{2\bm{k}\sigma}b^{}_{1\bm{k}\sigma}\right)\,.
\end{eqnarray}
Hence, for each value of the wave-number $\bm{k}$, we have a
$4\times4$ Hamiltonian $H_{\bm{k}}$ given by
\begin{equation}
H_{\bm{k}}=\left(\begin{array}{cccc} 
\varepsilon+V & t\phi({\bm{k}}) &  t_{4}\phi^{*}({\bm{k}}) & t_{\perp} \\  
t\phi^{*}({\bm{k}}) & -\varepsilon+V&  t_{3} \phi({\bm{k}}) &  t_{4}\phi^{*}({\bm{k}})\\
t_{4}\phi({\bm{k}}) &  t_{3}\phi^{*}({\bm{k}})& -\varepsilon-V &  t\phi({\bm{k}}) \\  
t_{\perp} & t_{4}\phi({\bm{k}})&  t\phi^{*}({\bm{k}}) & \varepsilon-V
 \end{array} \right)\,,
\label{eq:ham4x4}
\end{equation}
with eigenvectors
$[u^{n}_{A1}({\bm{k}}),u^{n}_{B1}({\bm{k}}),u^{n}_{A2}({\bm{k}}),u^{n}_{B2}({\bm{k}})]^{T}$
and eigenvalues $E_{n}({\bm{k}})$ with $n=1,2,3,4$. This Hamiltonian,
including $t_4$ has been written before \cite{konschuh_theory_2012} to
study spin-orbit effects of BLG at zero pressure.

\begin{figure}[t]
\includegraphics[width=\columnwidth]{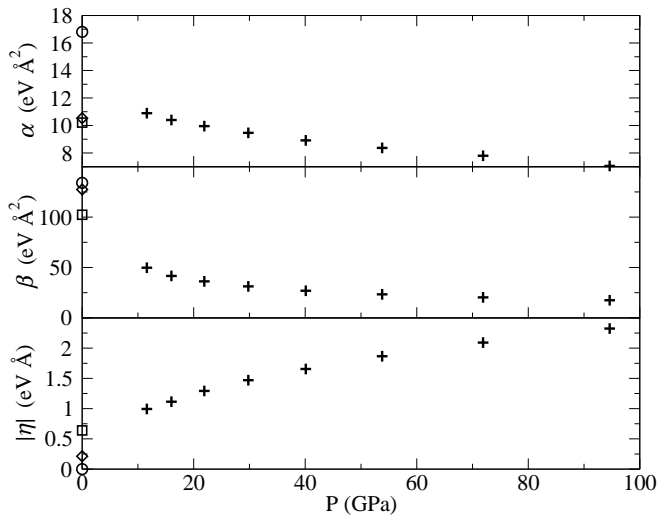}
\caption{Tight-binding parameters as function of pressure. (Plus
  signs) Values obtained by fitting the DFT-calculated bands along the
  T line of the Brillouin zone close to the point
  $\bm{K}$. ($\bigcirc$) Experimental values from
  Ref.~[\onlinecite{Zou11}].  ($\Box$) Experimental values from
  Ref.~[\onlinecite{zhang_determination_2008}]. ($\Diamond$)
  Experimental values from Ref.~[\onlinecite{Malard07}].}
\label{fig:tbpar}
\end{figure}

The low energy excitations with crystal momentum around the K and
K$^{\prime}$ points of the BZ can be described by an effective two
band Hamiltonian. The latter is obtained by eliminating the bands that
are shifted from the Fermi energy by $t_{\perp}$. As pressure
increases $t_{\perp}$ also increases improving the range of validity
of the approximation. Since we are interesting in describing the bands
near the K point, we can measure ${\bm{k}}$ from
${\bm{K}}=\frac{4\pi}{3\sqrt{3}a}(1,0)$ so that
$\phi(\bm{k})\simeq-\frac{3a}{2}\left(k_{x}-ik_{y}\right)$ for
$k=|\bm{k}|\ll|\bm{K}|$.  In the base of the $A_{2}$ and $B_{1}$
orbitals, the effective Hamiltonian takes the form
\begin{equation}
\label{Hblg}
H_{\bm{k}}^{\mathrm{eff}}=e(k) I+\bm{h}(\bm{k}) \cdot \bm{ \sigma}\,,
\end{equation}
where $ I$ is the unit matrix,
$\bm{\sigma}=(\sigma_x,\sigma_y,\sigma_z)$ with $\sigma_i$ are the
Pauli matrices,
\begin{equation}
\label{Hblge}
e(k)=-\varepsilon+\alpha k^2
\end{equation}
and $\bm{h}=(h_x,h_y,h_z)$ with 
\begin{eqnarray}
\nonumber
h_{x}(\bm{k})&=&-\eta\, k_{x}+\beta\left(k_{x}^{2}-k_{y}^{2}\right),\\
\nonumber
h_{y}(\bm{k})&=&\eta\, k_{y}+2\beta\, k_{x}k_{y},\\
h_{z}(\bm{k})&=&V\,.
\label{Hblgh}
\end{eqnarray}
Here $\eta=3t_{3}a/2$ and $\alpha$ and $\beta$ are functions of
microscopic parameters $t,t_{\perp},t_{3}$ and $t_{4}.$ In the lowest
order in $V$ and $\varepsilon$ we obtain
\begin{equation}
\alpha=\frac{9a^{2}}{2}\frac{tt_{4}}{t_{\perp}}
\end{equation}
and
\begin{equation}
\beta=\frac{9a^{2}}{4}\frac{t^{2}+t_{4}^{2}}{t_{\perp}}.
\end{equation}
%
\begin{table}[b]
  \caption{Tight binding parameters for the effective two-band model
    of Eq.~(\ref{Hblg}) obtained by fitting the bands of the DFT
    calculations. We include experimental values at normal pressure
    for comparison.
    \label{tbpar}
  }
  \begin{ruledtabular}
    \begin{tabular}{cccc}
      P (GPa)    & $\alpha$ (eV\AA$^2$) & $\beta$ (eV \AA$^2$)& $\eta$ (eV \AA) \\ \hline
      $\sim$0.1   & 16.81\footnotemark[1] & 133.97\footnotemark[1] & 0.00\footnotemark[1]\\
      $\sim$0.1   & 10.21\footnotemark[2] & 102.33\footnotemark[2] & 0.64\footnotemark[2]\\
      $\sim$0.1   & 10.53\footnotemark[3] & 127.40\footnotemark[3] & 0.21\footnotemark[3]\\
      11.6 & 10.895   & 49.815 & -0.995\\ 
      16.0 & 10.397   & 41.633 & -1.114\\ 
      21.9 &  9.950   & 36.236 & -1.293\\ 
      29.8 &  9.464   & 31.258 & -1.470\\ 
      40.1 &  8.917   & 26.922 & -1.655\\ 
      53.8 &  8.369   & 23.394 & -1.866\\ 
      71.9 &  7.796   & 20.375 & -2.092\\ 
      96.4 &  7.054   & 17.487 & -2.324\\ 
    \end{tabular}
  \end{ruledtabular}
 \footnotetext[1]{Experimental values from Ref.~[\onlinecite{Zou11}].}
 \footnotetext[2]{Experimental values from Ref.~[\onlinecite{zhang_determination_2008}].}
 \footnotetext[3]{Experimental values from Ref.~[\onlinecite{Malard07}].}
\end{table}
We have obtained the parameters of this effective two-band Hamiltonian
by fitting the bands of the DFT calculations in the region around the
K point. The results are provided in Table~\ref{tbpar} and the
resulting tight-binding bands are compared with the DFT-obtained bands
in Fig.~\ref{fig:BandsFull}. Also in the table, we quote the results
of measurements of the band structure of the BLG at normal
pressure. All values of the parameters given in Table~\ref{tbpar} are
plotted in Fig.~\ref{fig:tbpar} to show that the extrapolation of our
theoretical values towards zero pressure falls well within the
dispersion of the experimental values reported in the literature.  We
do not show calculated values at normal pressure because in his case,
the inclusion of van der Waals corrections are very important. For
higher pressures, the energy is dominated by the Pauli repulsion
between planes and the absence of this corrections does not introduce
any significant error.

Given this effective Hamiltonian, we can obtain the location of the
four Dirac cones by finding the points in the BZ where $h_x=h_y=0$
(assuming $V$ is the smallest energy scale). One of the Dirac cones is
always at $\bm{k}=(0,0)$ and the other three are at $\bm{k}=\bm{T}_i$
with $i=1,2,3$ and
\begin{eqnarray}
\bm{T}_1&=&\left(\frac{\eta}{\beta},0\right),\nonumber\\
\bm{T}_2&=&\frac{\eta}{\beta}\left(-\frac{1}{2},\frac{\sqrt{3}}{2}\right),\\
\bm{T}_3&=&\frac{\eta}{\beta}\left(-\frac{1}{2},-\frac{\sqrt{3}}{2}\right).\nonumber
\end{eqnarray}
By linearizing the effective Hamiltonian $H_{\bm{k}}^{\mathrm{eff}}$
for a small separation $\bm{q}$ from these points we reveal the band
structure at these Dirac cones and obtain expressions that facilitate
the evaluation of its topological properties and their
consequences. Around $\bm{k}=(0,0)$ (K point) we then have
\begin{equation}
h_x(\bm{q})\simeq-\eta q_x\,,\qquad
h_y(\bm{q})\simeq\eta q_y\,, \qquad
e(\bm{q})\simeq-\varepsilon\,,
\label{hK}
\end{equation}
which gives the following  eigenenergies
\begin{equation}
\label{eneK}
E_{\bm{K}}^{\pm}(\bm{q})=-\varepsilon\pm\sqrt{V^2+\eta^2\left(q_x^2+q_y^2\right)}.
\end{equation}
Similarly,  around $\bm{T}_1$ we find that 
\begin{eqnarray}
\label{hT}
h_x(\bm{T}_1+\bm{q})&\simeq&\eta q_x,\nonumber\\
h_y(\bm{T}_1+\bm{q})&\simeq&3\eta q_y, \\
e(\bm{T}_1+\bm{q})&\simeq&-\varepsilon+\alpha\frac{\eta^2}{\beta^2}+2\alpha\frac{\eta}{\beta}q_x,\nonumber
\end{eqnarray}
and the corresponding eigenenergies
\begin{equation}
\label{eneT}
E_{\bm{T}_1}^{\pm}(\bm{q})=-\varepsilon+\alpha\frac{\eta^2}{\beta^2}+2\alpha\frac{\eta}{\beta}q_x\pm\sqrt{V^2+\eta^2\left(q_x^2+9q_y^2\right)}.
\end{equation}
Expressions for $\bm{T}_2$ and $\bm{T}_3$ are similar and can be
obtained by symmetry considerations or by direct calculation.

\begin{figure}[ht]
\includegraphics[width=\columnwidth]{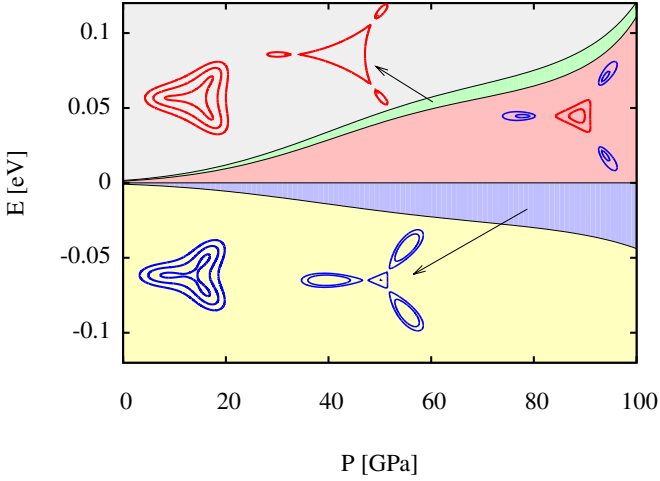}
\caption{(Color online) The (E,P) plane with the four lines defining
  regions with different topology of the Fermi lines, in each region
  the form of constant energy curves are shown, blue and red lines
  indicate the portions corresponding the valence and conduction bands
  respectively.}
\label{fig:ep-plane}
\end{figure}

It is clear from Fig.~\ref{fig:BandsFull} that there are four
cha\-rac\-te\-ris\-tic energies that define the low energy band
structure (taking $V=0$): $E_{\bm{K}}^{\pm}(\bm{0})=-\varepsilon$,
corresponding to the apex of the Dirac cone centred at K, the energy
$E_{\bm{T}_1}^{\pm}(\bm{0})$ of the apex of the cones shifted from K
and the energies $E_{\mathrm{sv}}$ and $E_{\mathrm{sc}}$ of the saddle
points (relative minimum and maximum) of the valence and conduction
bands, respectively. These four energies divide the energy-pressure
plane in five different regions (see Fig.~\ref{fig:ep-plane}). In each
of them the surface of constant energy consists of different pieces
with different topological properties. For neutral BLG, the Fermi
energy is larger than $E_{\bm{K}}^{\pm}(\bm{0})$ and smaller than
$E_{\bm{T}_1}^{\pm}(\bm{0})$ and the 2D Fermi surface consists of an
electron pocket centred at $\bm{K}$ and three hole pockets centred at
$\bm{T}_1$ and its symmetrically equivalent points. The same occurs
around the K$^{\prime}$ point of the BZ.

Comparing the band energies at $\bm{K}$ and at $\bm{T}_1$ we obtain 
\begin{equation}
E_{\bm{T}_1}^\pm(\bm{0})-E_{\bm{K}}^\pm(\bm{0})=\alpha\frac{\eta^2}{\beta^2}\propto t_4\,,
\end{equation}
which is proportional to $t_4$. This lead us to an important point:
\textit{without including this term in the tight binding model, the
  trigonal warping does not separate the cones in energy and leads to
  the wrong conclusion that their individual properties would not
  manifested in transport or spectroscopic experiments}. This is not
the case as we show in the following sections.

\section{Berry Curvature and Valley Hall Effect}
\label{sec:berry}
The interesting band topology of BLG unveiled by the effect of
pressure offers a unique possibility to study valley polarization and
the valley Hall effect \cite{xiao_valley-contrasting_2007}. In
contrast to the graphene case, this system only requires to apply a
small electric field perpendicular to the bilayer to reveal this
phenomena. In this section we will use our previously developed
tight-binding model to evaluate the Berry curvature of the bands and
the transverse conductivity as a function of pressure.

With the four band Hamiltonian of Eq.~(\ref{eq:ham4x4}), the Berry
curvature of the $n$-th band is given by
\begin{equation}
\Omega_{z}^{n}=2\,\mathrm{Im}\sum_{n^{\prime}\ne
  n}\frac{\langle n|\frac{\partial H_{\bm{k}}}{\partial
    k_{x}}|n^{\prime}\rangle\langle n^{\prime}|\frac{\partial H_{\bm{k}}}{\partial
    k_{y}}|n\rangle}{(E_{n}({\bm{k}})-E_{n^{\prime}}({\bm{k}}))^{2}}.
\end{equation}
This expression, however, requires a simple but nonetheless
unnecessary numerical work.
Since the Berry curvature is concentrated around the location of the
Dirac cones, we can use the two band approximation of
Eqs.~(\ref{Hblg}), (\ref{Hblge}), and (\ref{Hblgh}) to calculate the
Berry curvature of these bands.\footnote{We have compared the result
  of calculating the Berry curvature from the $4\times4$ model and the
  $2\times2$ approximation and the difference is negligible.}  In this
case~\cite{Hasan2010}
\begin{equation}
\label{omega2x2}
\Omega^\pm_{z}=\mp\frac{1}{2h^3}\bm{h}\cdot\partial_{k_{x}}\bm{h}\times\partial_{k_{y}}\bm{h}
\end{equation}
where $h^3=\left(h_{x}^2+h_{y}^2+h_{z}^2\right)^{3/2}$. This finally gives
\begin{equation}
\label{omegaall}
\Omega^\pm_{z}={\frac{\mp V\left[4\beta^2k^2-\eta^2\right]} 
{2\left[\beta^2k^4+\eta k^2+2\eta\beta k_x\left( 2 k_y^2 -k_x^2\right)+V^2\right]^{3/2}}},
\end{equation}
where $k^2=k_x^2+k_y^2$.  The intrinsic contribution to the anomalous
Hall effect is given by \cite{xiao_berry_2010}
\begin{equation}
\sigma_{xy}=\frac{e^{2}}{\hbar}\int\frac{d\bm{k}}{\left(2\pi\right)^{2}}\sum_{s=\pm}f\left(E_{s}\left(\bm{k}\right)\right)\Omega^s_{z}\left(\bm{k}\right).
\end{equation}
Given that the curvature is substantially large around the conical
intersections compared with its value around other regions of the BZ,
we can use the approximated expressions around $\bm{K}$ and
$\bm{T}_{1}$ to simplify this expression. To a good degree of
approximation this conductivity can be calculated as
\begin{equation}
\sigma_{xy}=\sigma_{\bm{K}}+\sum_{i=1}^{3}\sigma_{\bm{T}_{i}}=\sigma_{\bm{K}}+3\sigma_{\bm{T}_{1}}
\end{equation}
where we have used the symmetry of the three $\bm{T}_{i}$ points and
have defined
\begin{equation}
\sigma_{\bm{K}}=\frac{e^{2}}{\hbar}\int\frac{d\bm{q}}{\left(2\pi\right)^{2}}\sum_{s=\pm}f\left(E^s_{\bm{K}}\left(\bm{q}\right)\right)\Omega^s_{z}\left(\bm{q}\right)
\end{equation}
and
\begin{equation}
\sigma_{\bm{T}_{1}}=\frac{e^{2}}{\hbar}\int\frac{d\bm{q}}{\left(2\pi\right)^{2}}\sum_{s=\pm}f\left(E^{s}_{\bm{T}_1}\left(\bm{q}\right)\right)\Omega^s_{z}\left(\bm{T}_{1}+\bm{q}\right)\,.
\end{equation}
The integration around the $\bm{K}$ point can be done a\-na\-ly\-ti\-ca\-lly at zero temperature using the approximation (obtained from Eq. (\ref{omegaall}))
\begin{equation}
\Omega_z^{\pm}\left(\bm{q}\right)\simeq\pm\frac{V\eta^{2}}{2\left(V^{2}+\eta^{2}q^{2}\right)^{3/2}}\,,
\end{equation}
so that its contribution to the conductivity results 
\begin{equation}
\sigma_{\bm{K}}=\begin{cases}
\begin{array}{cc}
\frac{e^{2}}{2h}\frac{V}{\mu+\varepsilon} & \mu<-\varepsilon-V\\
-\frac{e^{2}}{2h} & -\varepsilon-V<\mu<-\varepsilon+V\\
-\frac{e^{2}}{2h}\frac{V}{\mu+\varepsilon} & \mu>-\varepsilon+V
\end{array}\end{cases}\,.
\end{equation}
where $\mu$ is the chemical potential.
Around $\bm{T}_{1}$ we can use that
\begin{equation}
\Omega_z{\pm}\left(\bm{T}_{1} + \bm{q}\right)\simeq
\mp\frac{3V\eta^{2}}{2\left(V^{2} + \eta^{2}q_{x}^{2} +
  9\eta^{2}q_{y}^{2}\right)^{3/2}}\,.
\end{equation}
In this case, however, the presence of $q_{x}$ outside the square root
in the energy, Eq.~(\ref{eneT}), makes the integration rather
difficult. Nevertheless, the contribution to the conductivity from
this region of the BZ can be approximated as
\begin{equation}
\sigma_{\bm{T}_{1}}=\begin{cases}
\begin{array}{cc}
-\frac{e^{2}}{2h}\frac{V}{\mu-E_{0}} & \mu<E_{0}-V\\
\frac{e^{2}}{2h} & E_{0}-V<\mu<E_{0}+V\\
+\frac{e^{2}}{2h}\frac{V}{\mu-E_{0}} & \mu>E_{0}+V
\end{array}\end{cases}\,,
\end{equation}
where $E_{0}=-\varepsilon+\alpha\frac{\eta^{2}}{\beta^{2}}$ is the
center of the band gap opened by the perpendicular electric field.

\begin{figure}
\includegraphics[width=\columnwidth]{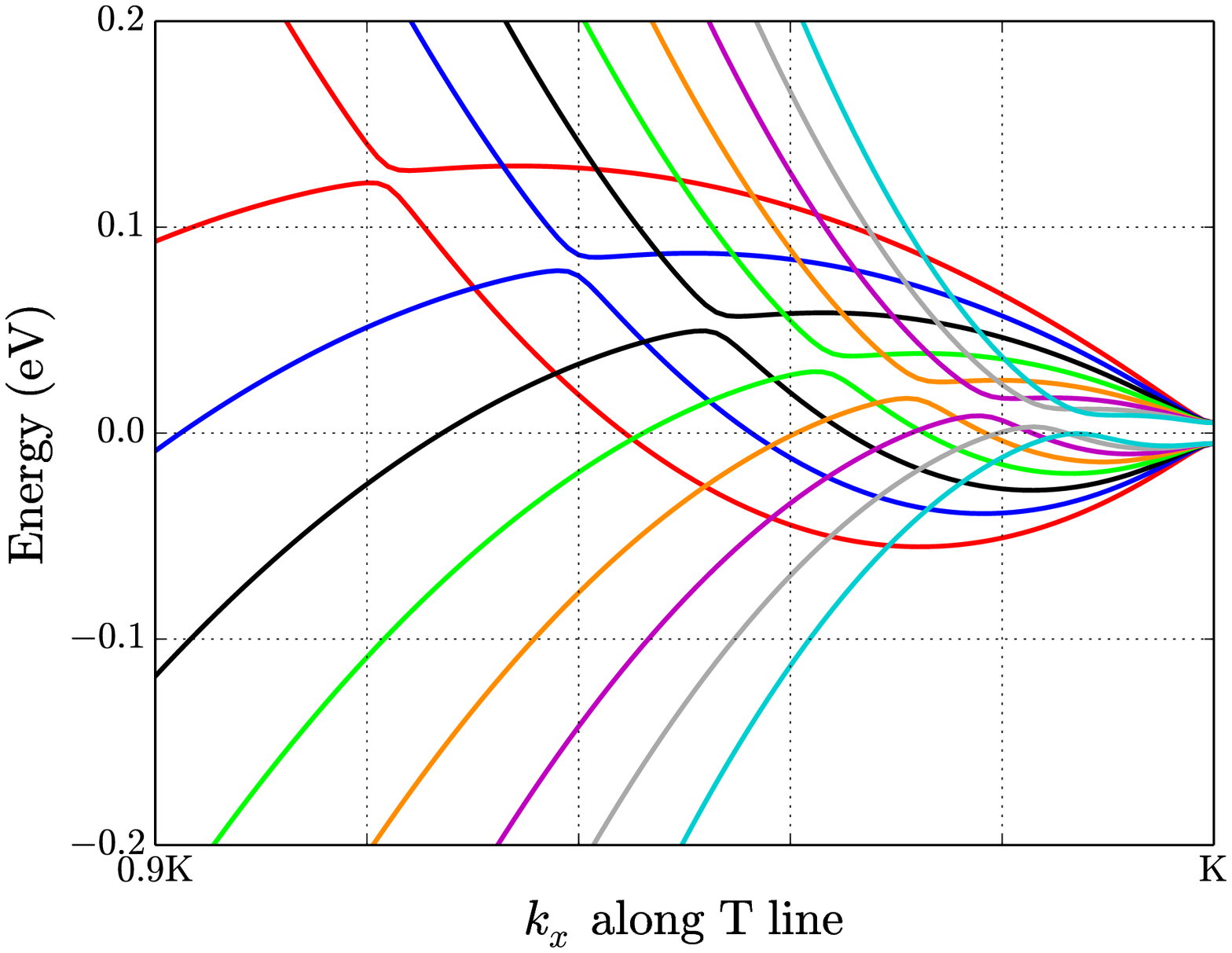}
\includegraphics[width=\columnwidth]{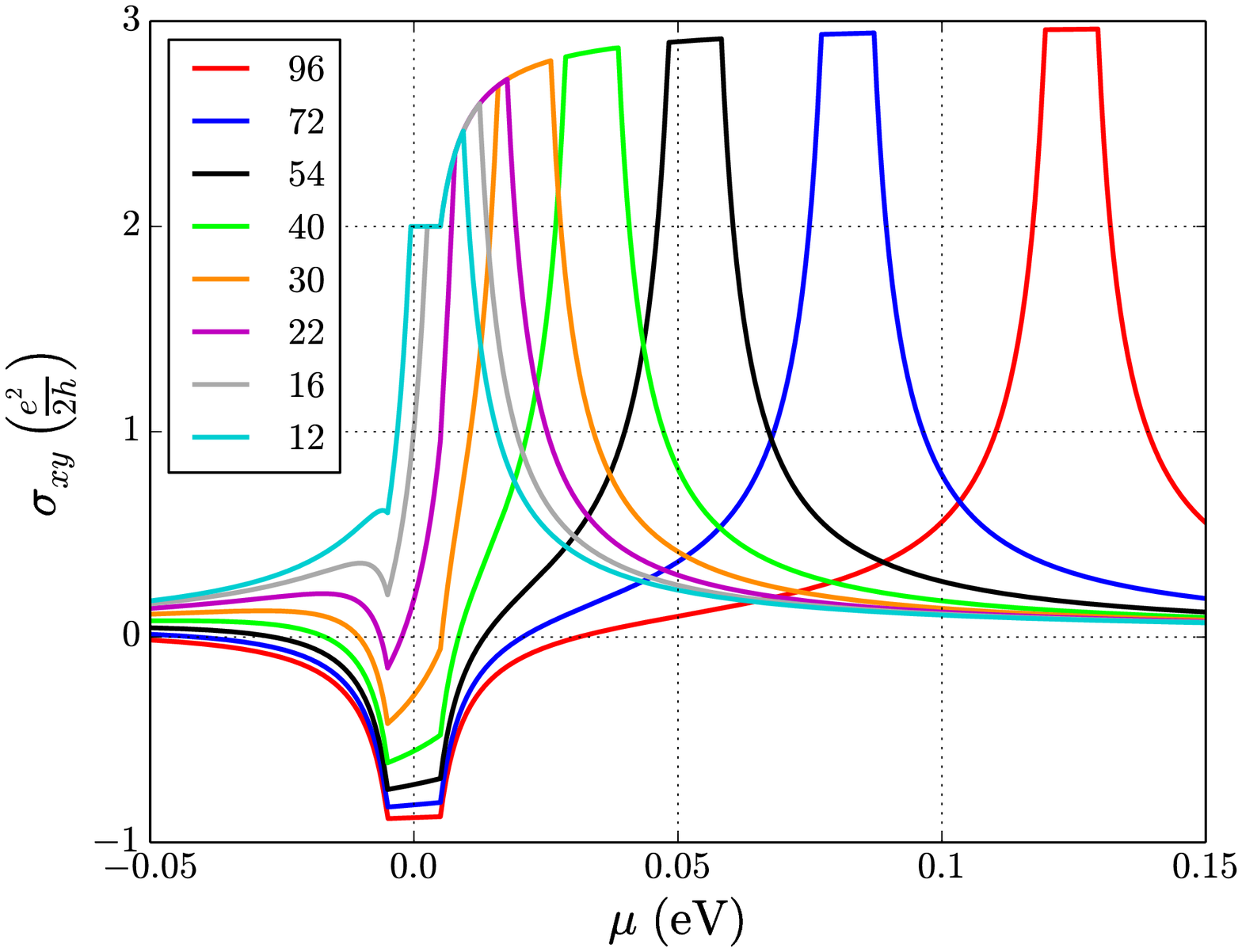}
\caption{(Upper panel) Energy bands of the bilayer at different
  pressures under the effect of an electric field perpendicular to the
  bilayer equivalent to $V=0.005$~eV. (Bottom panel) Intrinsic
  contribution to the anomalous Hall conductivity as function of the
  chemical potential for different applied pressures and the same
  applied field.}
\label{sigmaP}
\end{figure}

We can now calculate the expected variation with applied pressure of
the total intrinsic contribution to the anomalous Hall conductivity as
a function of the chemical potential as shown in Fig.~\ref{sigmaP}.
In this figure we see that with increasing pressure the contribution
from the cones at the $\bm{T}_{i}$ points moves to higher energies and
produces an anomalous Hall conductivity that is three times larger,
and of opposite sign, than the contribution (observed at lower $\mu$)
produced by the cone at $\bm{K}$.  Close to zero pressure, our
approximations are bound to be worst because the Dirac cones at
$\bm{T}_{1}$ are closer to $\bm{K}$. However, in this figure it is
clear that the anomalous conductivity as the pressure becomes smaller
tends the values for BLG at zero
pressure.\cite{novoselov_unconventional_2006} This dramatic change
with pressure, made possible by the electron-hole symmetry breaking,
can have interesting applications as a very sensitive pressure gauge.
Apart from the already interesting phenomena of the direct observation
of a valley hole effect in BLG, it is important to highlight that the
signature of this effect will be the ratio of $3$ to $1$ as the
chemical potential is swipes though the gaps at $\bm{K}$ and
$\bm{T}_1$.

\section{Landau Levels}
The LLs in BLG have received much attention during the last
decade.\cite{yin_experimental_2015,kawarabayashi_stability_2013} The
simplest description in the absence of electric fields predicts a
spectrum with two-fold zero energy states (per spin) for each one of
the Dirac cones at the K and the K$^{\prime}$ points of the BZ. The
trigonal warping, given by the parameter $\beta$ (that preserves the
electron-hole symmetry), strongly modifies the spectrum at low
fields. In fact, as we showed in the previous sections, this
perturbation modifies the parabolic bands around the Dirac points
leading to four degenerate cones in each corner of the BZ. As a
consequence, for low magnetic fields the zero energy states at K
(K$^{\prime}$) have a four-fold degeneracy (per spin).  Pressure
breaks electron-hole symmetry shifting the energy of the cones and
removes some of the degeneracies of the low energy LLs.  Moreover, as
we show below, all states have a field dependent energy. The
robustness of the zero energy LL characteristic of monolayer and
multilayer graphene is lost in the high pressure regime.

To calculate the Landau Levels of BLG under pressure we proceed as in
Refs. [\onlinecite{de_gail_magnetic_2012}] and
[\onlinecite{yuan_landau_2011}]. In the continuous limit and in the
presence of an external magnetic field $\bm{B}$ the canonical momentum
$\bm{p}$ must be replaced by
$\bm{\Pi}=\bm{p}+e\bm{A}\left(\bm{r}\right)$ where
$\bm{A}\left(\bm{r}\right)$ is the vector potential describing a
magnetic field perpendicular to the graphene layers. We use units such
that $\hbar\equiv1\equiv c.$ The components of the gauge-invariant
momentum obey the commutation relation
$\left[\Pi_{x},\Pi_{y}\right]=-i/l_{B}^{2}$ where
$l_{B}\simeq26$nm$/\sqrt{B\left[\mathrm{T}\right]}$ is the magnetic
length.  This allows us to introduce the harmonic oscillator operators
$\hat{a}=\lambda^{-1}\Pi_{-}$ and
$\hat{a}^{\dagger}=\lambda^{-1}\Pi_{+}$ with
$\left[\hat{a},\hat{a}^{\dagger}\right]=1$,
$\lambda=\left(l_{B}/\sqrt{2}\right)^{-1}$ and where
$\Pi_{\pm}=\Pi_{x}\pm i\Pi_{y}.$ In what follows we use the Landau
gauge $\bm{A\left(\bm{r}\right)}=\left(0,Bx\right)$.
Then, the effective two band Hamiltonian takes the form
\begin{equation}
H=\left(\begin{array}{cc}
-\varepsilon+\alpha\lambda^{2}\hat{a}^{\dagger}\hat{a} & -\eta\lambda\hat{a}+\beta\lambda^{2}\hat{a}^{\dagger2}\\
-\eta\lambda\hat{a}^{\dagger}+\beta\lambda^{2}\hat{a}^{2} & -\varepsilon+\alpha\lambda^{2}\hat{a}\hat{a}^{\dagger}
\end{array}\right)\,.
\label{hlls}
\end{equation}
As usual, in the base of Landau functions
$\psi\left(x,y\right)=e^{iky}\varphi_{n}(x)$ the harmonic oscillator
operator satisfy
\begin{eqnarray}
\nonumber
\hat{a}^{\dagger}\varphi_{n}&=&\sqrt{n+1}\,\varphi_{n+1}\\
\hat{a}\varphi_{n}&=&\sqrt{n}\,\varphi_{n-1}\,.
\end{eqnarray}
With the minimum inter-plane coupling, $\alpha=\eta=0$ and the charge
transfer energy $\varepsilon=0$, the model reduces to the simplest
model used to describe BLG at room pressure. In this case the LLs
spectrum contains a zero energy doublet and states with energies
${E_{n}}^{\pm}=\pm \sqrt{n(n-1)} \hbar \omega_{c}$ with $n\geq 2$ and
$ \hbar \omega_{c}=\beta \lambda^{2}$. The corresponding wavefunctions
are
\begin{equation}
\chi_{0}=\left[ \begin{array}{c} \varphi_{0}\\ 0 \end{array} \right]\,,\qquad
\chi_{1}=\left[ \begin{array}{c} \varphi_{1}\\ 0\end{array} \right]\,,\qquad
    {\chi_{n}}^{\pm}=\left[ \begin{array}{c}\varphi_{n}\\{\pm}
        \varphi_{n-2}\end{array} \right]\,.
\end{equation}
 \begin{figure}[tb]
\includegraphics[width=\columnwidth]{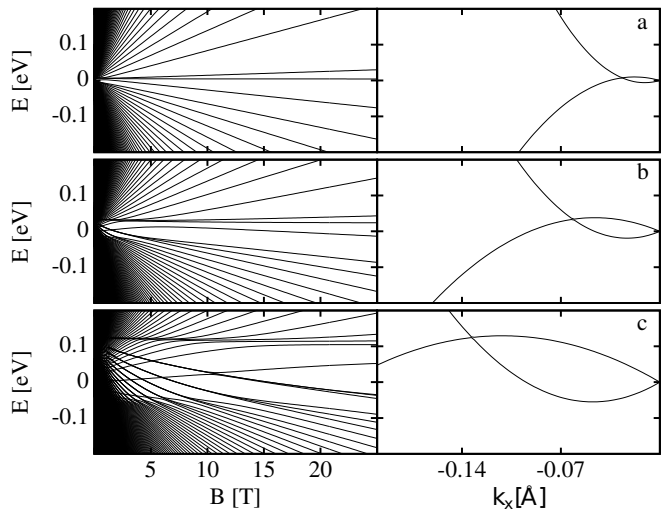}
\caption{Left and right panels correspond to the LLs spectrum and the
  zero field band structure, respectively, for different values of
  pressure: (a) $16$ GPa; (b) $40$ GPa; (c) $96$ GPa.}
\label{fig_landau_levels_pressure}
\end{figure}  
For small pressure the LLs energies are shifted by a first order
correction due to the $\alpha$-terms in Hamiltonian
(\ref{hlls}). These corrections break the electron-hole symmetry and
lift the degeneracy of the zero energy modes. The first-order
corrected spectrum has $E_{0}=0$, $ E_{1}=\alpha \lambda^{2}$ and, for
$n\geq 2$, $E_{n}=\pm \sqrt{n(n-1)} \hbar \omega_{c}+(2n-1)\alpha
\lambda^{2}$. This effect can be observed in
Fig.~\ref{fig_landau_levels_pressure}(a).  Further second order
corrections due to the $\eta$ and $\alpha$ terms were also
evaluated. However, as the pressure increases, perturbation theory is
not enough to account for the evolution of the LLs spectrum. The
numerically obtained spectrum for different pressures is shown in the
left panels of Fig.~\ref{fig_landau_levels_pressure}.

For high pressures and low energies
($E_{\mathrm{sv}}<E<E_{\mathrm{sc}}$) we can get some insight into the
structure of the LLs spectrum by assuming the presence of four
independent Dirac cones, centered at the $\bm{K}$ and $\bm{T}_{i}$
points. The corresponding spectrum is given by
$E_{\bm{K},n}=E_{\bm{K},0} \pm C_{1} \sqrt{Bn}$ and
$E_{\bm{T}_1,n}=E_{\bm{T}_1,0}\pm C_{2} \sqrt{Bn}$ where $C_{1}$ and
$C_{2}$ are constants.
\begin{figure}[b]
\includegraphics[width=\columnwidth]{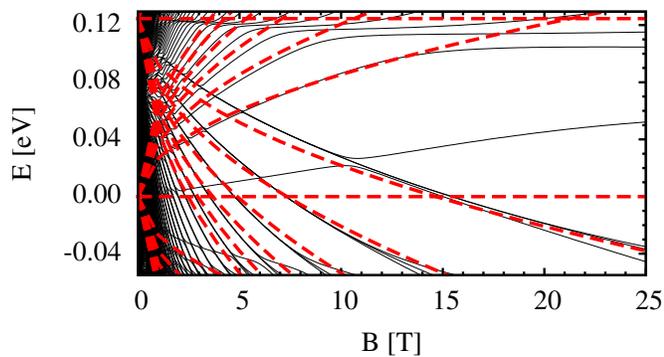}
\caption{(Color online) LLs spectrum obtained numerically for $96$ GPa
  (continuous black lines) and LLs spectrum corresponding to two
  independent Dirac cone centered at $\bm{K}$ and $\bm{T}_{1}$ (dashed
  red lines).}
\label{fig5}
\end{figure}  
While the structure of the so obtained LLs spectrum is similar to the
numerically obtained results, there are important differences (see
Fig.~\ref{fig5}). First, in the numerical results, it is evident that
the energies of the $n=0$ LL states of the four cones are no longer
field independent---quantum corrections for the low $n$ states are
important even for small fields. In addition, there are anticrossings
that cannot be captured by in a picture that treats the cones
independently. We also notice that as the energy of the upper cone LLs
approach the saddle point energy $E_{\mathrm{sv}}$, the three fold
degeneracy is lifted. These last effects are due to the magnetic
breakdown, \textit{i.e.} Landau-Zener tunneling mixing the states of
the different cones.

A similar and more qualitative analysis can be done by resorting to
the Onsager's semiclassical quantization rule
\cite{onsager_interpretation_1952}. This rule sates that the area
enclosed by orbits in $k$-space are quantized according to the
following condition \cite{fuchs_topological_2010}
\begin{equation}
A(E_{m})=2\pi e B (m+\gamma)\,.
\label{onsag}
\end{equation}
Here $E_{m}$ is the energy of the orbit and $\gamma$ is a constant
($0\leq \gamma < 1$) independent of the quantum number $m$. Onsager's
semiclassical approach is well justified for large $m$ only. However,
for the sake of comparison, we shall use it without restriction,
including the $m=0$ states. Concerning the quantization condition of
Eq. (\ref{onsag}) it was recently shown that, in two band materials
like the one under consideration, the constant $\gamma$ is given by
the pseudo-spin winding number $w_{C}$ which is $1/\pi$ times the
Berry phase obtained along the close orbit
\cite{fuchs_topological_2010}
\begin{equation}
\gamma=\frac{1}{2}-\frac{|w_{C}|}{2}\,.
\end{equation}
\begin{figure}[tb]
  \includegraphics[width=\columnwidth]{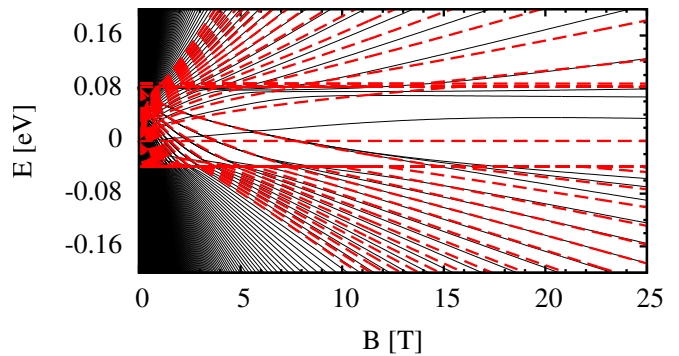}
  \caption{(Color online) Comparison between LLs spectrum obtained
    numerically (continuous black lines) and using Onsager's
    quantization (dashed red lines), see Eq. (\ref{onsag}), under $72$
    GPa.}
  \label{fig6}
\end{figure}
After evaluation the areas as a function of the energy and calculating
the winding numbers corresponding to the different constant energy
close-orbits, we invert Eq. (\ref{onsag}) to obtain the semiclassical
spectrum. The comparison of the semiclassical and the numerical
results shown in Fig.~\ref{fig6} is quite good for not too small
values of $n$. The LLs corresponding to small $n$ do not
follow Onsager's rule. Moreover, the numerical results show that all
LLs have a field dependent energy showing that in the presence of the
electron-hole symmetry breaking parameter $\alpha$ the stability of
the zero-modes LL is lost.

\begin{figure}[ht!]
\includegraphics[width=\columnwidth]{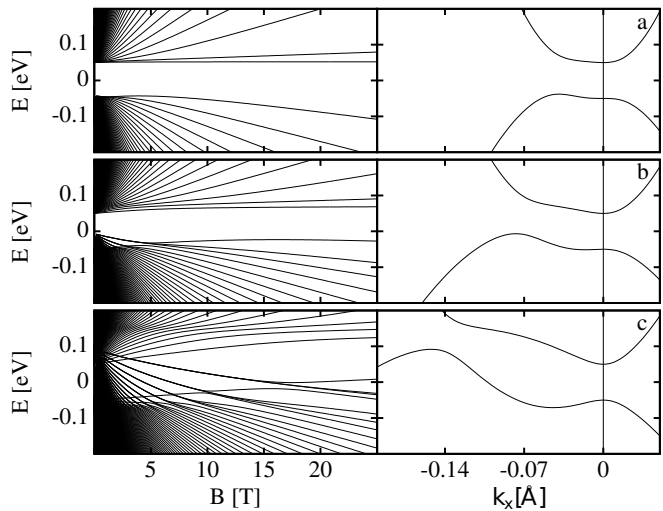}
\caption{The same that Fig.~\ref{fig_landau_levels_pressure} but in
  the presence of a perpendicular electric field $V=50$meV.}
\label{fig7}
\end{figure}

Finally, the LLs spectrum in the presence of a perpendicular electric
field $V=50$meV is shown in Fig.~\ref{fig7} for different
pressures. With the minimum inter-layer coupling ($t_3=t_4=0$), any
electric field opens a gap and the LLs spectrum displays such a gap
with a set of LLs associated to the conduction band, whose energies
increase with increasing magnetic field, separated from those
corresponding to the valence band with opposite slope. As the pressure
increases the gap decreases and for a given (electric field dependent)
critical pressure the gap closes as a consequence of the band crossing
as discussed in the previous section.

\section{Summary and Conclusions}
We have shown that upon the application of pressure the low energy
band structure of BLG undergoes an interesting transformation. Two
parabolic bands touching at the $\bm{K}$ point of the Brillouin zone
with phase winding 2$\pi$ transform into a set of four Dirac cones,
one with a winding number of $\pi$ that remains at $\bm{K}$ and three
that separate along the $\bm{T}$ lines with a winding number of
-$\pi$. The most interesting discovery is that the cones at $\bm{T}$
separate in energy from the cone at $\bm{K}$ generating a Lifshitz
transition visible by experiment.  The study of the electronic
structure using DFT and the interpretation of the results in terms of
a tight-binding model show that the main effects of pressure are the
following: (i) an increase of the direct $A_{1}B_{2}$ hopping
integrals $t_{\perp}$. As a result, two bands are shifted away from
the Dirac point improving the description of the system in terms of an
effective two band model; (ii) an increase of the $t_{3}$ and $t_{4}$
hopping integrals. These parameters modify the low energy band
structure generating four Dirac cones at the two corners of the BZ. An
important effect is the breaking of the electron-hole symmetry; and
(iii) a small charge transfer between the two non-equivalent sites of
each plane, an effect that is not relevant when describing the low
energy bands in terms of a simple two band model.

The resulting band structure is summarized in Fig.~\ref{fig:ep-plane}
where the energy-pressure plane is divided in five regions with
different constant energy surfaces.  Our estimates of the pressure
dependence of the microscopic parameters are in good agreement with
the zero pressure extrapolation as obtained in recent experimental
works.

The evolution of the band structure with pressure can be measured by
means of different experimental techniques. Although the small
buckling of the graphene planes with pressure does not produce a
significant increase of the spin-orbit coupling, valley Hall effect
and the structure of the Landau levels show important changes
unveiling the pressure induced Lifshitz transition.

We presented results of the Hall conductivity and show that this
quantity is very sensitive to the pressure and to the carrier
density. Variations in the Fermi level change the sign and the
magnitude of the valley Hall response that is increased by a factor
three as the carriers vary their origin from the single Dirac cone at
$\bm{K}$ to the triple degenerate Dirac cones at $\bm{T}$. The high
sensitivity of this effect with pressure makes this structure a
potential source for the design of a device for the detection of small
pressure variations on two dimensional surfaces such as screen devices
and joints.

The LLs evolve from a simple structure at low pressures to a rich
spectrum at high pressures. The most notable effect is the breaking of
the electron-hole symmetry and the lack of stability of the zero-LL
modes, whereas all modes have a field dependent energy.

\begin{acknowledgments}
FM is supported by `Financiamiento Basal para Centros Cientificos y
Tecnologicos de Excelencia FB 0807' and Fondecyt grant \# 1150806. JOS
and CAB are grateful to the American Physical Society International
Travel Grant Awards Program that supported the visit of CAB to Penn
State when this work was started.  HPOC, GU and CAB acknowledge
financial support from PICTs 2013-1045 and Bicentenario 2010-1060 from
ANPCyT, PIP 11220110100832 from CONICET and grant 06/C415 from
SeCyT-UNC. GU acknowledges support from the ICTP associateship program
and the Simons Foundation.
\end{acknowledgments}


\begin{thebibliography}{53}%
\makeatletter
\providecommand \@ifxundefined [1]{%
 \@ifx{#1\undefined}
}%
\providecommand \@ifnum [1]{%
 \ifnum #1\expandafter \@firstoftwo
 \else \expandafter \@secondoftwo
 \fi
}%
\providecommand \@ifx [1]{%
 \ifx #1\expandafter \@firstoftwo
 \else \expandafter \@secondoftwo
 \fi
}%
\providecommand \natexlab [1]{#1}%
\providecommand \enquote  [1]{``#1''}%
\providecommand \bibnamefont  [1]{#1}%
\providecommand \bibfnamefont [1]{#1}%
\providecommand \citenamefont [1]{#1}%
\providecommand \href@noop [0]{\@secondoftwo}%
\providecommand \href [0]{\begingroup \@sanitize@url \@href}%
\providecommand \@href[1]{\@@startlink{#1}\@@href}%
\providecommand \@@href[1]{\endgroup#1\@@endlink}%
\providecommand \@sanitize@url [0]{\catcode `\\12\catcode `\$12\catcode
  `\&12\catcode `\#12\catcode `\^12\catcode `\_12\catcode `\%12\relax}%
\providecommand \@@startlink[1]{}%
\providecommand \@@endlink[0]{}%
\providecommand \url  [0]{\begingroup\@sanitize@url \@url }%
\providecommand \@url [1]{\endgroup\@href {#1}{\urlprefix }}%
\providecommand \urlprefix  [0]{URL }%
\providecommand \Eprint [0]{\href }%
\providecommand \doibase [0]{http://dx.doi.org/}%
\providecommand \selectlanguage [0]{\@gobble}%
\providecommand \bibinfo  [0]{\@secondoftwo}%
\providecommand \bibfield  [0]{\@secondoftwo}%
\providecommand \translation [1]{[#1]}%
\providecommand \BibitemOpen [0]{}%
\providecommand \bibitemStop [0]{}%
\providecommand \bibitemNoStop [0]{.\EOS\space}%
\providecommand \EOS [0]{\spacefactor3000\relax}%
\providecommand \BibitemShut  [1]{\csname bibitem#1\endcsname}%
\let\auto@bib@innerbib\@empty
\bibitem [{\citenamefont {Novoselov}\ \emph {et~al.}(2004)\citenamefont
  {Novoselov}, \citenamefont {Geim}, \citenamefont {Morozov}, \citenamefont
  {Jiang}, \citenamefont {Zhang}, \citenamefont {Dubonos}, \citenamefont
  {Grigorieva},\ and\ \citenamefont {Firsov}}]{novoselov_electric_2004}%
  \BibitemOpen
  \bibfield  {author} {\bibinfo {author} {\bibfnamefont {K.~S.}\ \bibnamefont
  {Novoselov}}, \bibinfo {author} {\bibfnamefont {A.~K.}\ \bibnamefont {Geim}},
  \bibinfo {author} {\bibfnamefont {S.~V.}\ \bibnamefont {Morozov}}, \bibinfo
  {author} {\bibfnamefont {D.}~\bibnamefont {Jiang}}, \bibinfo {author}
  {\bibfnamefont {Y.}~\bibnamefont {Zhang}}, \bibinfo {author} {\bibfnamefont
  {S.~V.}\ \bibnamefont {Dubonos}}, \bibinfo {author} {\bibfnamefont {I.~V.}\
  \bibnamefont {Grigorieva}}, \ and\ \bibinfo {author} {\bibfnamefont {A.~A.}\
  \bibnamefont {Firsov}},\ }\bibfield  {title} {\enquote {\bibinfo {title}
  {Electric {Field Effect} in {Atomically Thin Carbon Films}},}\ }\href
  {\doibase 10.1126/science.1102896} {\bibfield  {journal} {\bibinfo  {journal}
  {Science}\ }\textbf {\bibinfo {volume} {306}},\ \bibinfo {pages} {666}
  (\bibinfo {year} {2004})}\BibitemShut {NoStop}%
\bibitem [{\citenamefont {Zhang}\ \emph {et~al.}(2005)\citenamefont {Zhang},
  \citenamefont {Tan}, \citenamefont {Stormer},\ and\ \citenamefont
  {Kim}}]{Zhang2005}%
  \BibitemOpen
  \bibfield  {author} {\bibinfo {author} {\bibfnamefont {Y.}~\bibnamefont
  {Zhang}}, \bibinfo {author} {\bibfnamefont {Y.-W.}\ \bibnamefont {Tan}},
  \bibinfo {author} {\bibfnamefont {H.~L.}\ \bibnamefont {Stormer}}, \ and\
  \bibinfo {author} {\bibfnamefont {P.}~\bibnamefont {Kim}},\ }\bibfield
  {title} {\enquote {\bibinfo {title} {Experimental observation of the quantum
  hall effect and and berry's phase in graphene},}\ }\href {\doibase
  10.1038/nature04235} {\bibfield  {journal} {\bibinfo  {journal} {Nature}\
  }\textbf {\bibinfo {volume} {438}},\ \bibinfo {pages} {201} (\bibinfo {year}
  {2005})}\BibitemShut {NoStop}%
\bibitem [{\citenamefont {Castro~Neto}\ \emph {et~al.}(2009)\citenamefont
  {Castro~Neto}, \citenamefont {Guinea}, \citenamefont {Peres}, \citenamefont
  {Novoselov},\ and\ \citenamefont {Geim}}]{castro_neto_electronic_2009}%
  \BibitemOpen
  \bibfield  {author} {\bibinfo {author} {\bibfnamefont {A.~H.}\ \bibnamefont
  {Castro~Neto}}, \bibinfo {author} {\bibfnamefont {F.}~\bibnamefont {Guinea}},
  \bibinfo {author} {\bibfnamefont {N.~M.~R.}\ \bibnamefont {Peres}}, \bibinfo
  {author} {\bibfnamefont {K.~S.}\ \bibnamefont {Novoselov}}, \ and\ \bibinfo
  {author} {\bibfnamefont {A.~K.}\ \bibnamefont {Geim}},\ }\bibfield  {title}
  {\enquote {\bibinfo {title} {The electronic properties of graphene},}\ }\href
  {\doibase 10.1103/RevModPhys.81.109} {\bibfield  {journal} {\bibinfo
  {journal} {Rev. Mod. Phys.}\ }\textbf {\bibinfo {volume} {81}},\ \bibinfo
  {pages} {109} (\bibinfo {year} {2009})}\BibitemShut {NoStop}%
\bibitem [{\citenamefont {Balandin}\ \emph {et~al.}(2008)\citenamefont
  {Balandin}, \citenamefont {Ghosh}, \citenamefont {Bao}, \citenamefont
  {Calizo}, \citenamefont {Teweldebrhan}, \citenamefont {Miao},\ and\
  \citenamefont {Lau}}]{Balandin2008}%
  \BibitemOpen
  \bibfield  {author} {\bibinfo {author} {\bibfnamefont {A.~A.}\ \bibnamefont
  {Balandin}}, \bibinfo {author} {\bibfnamefont {S.}~\bibnamefont {Ghosh}},
  \bibinfo {author} {\bibfnamefont {W.}~\bibnamefont {Bao}}, \bibinfo {author}
  {\bibfnamefont {I.}~\bibnamefont {Calizo}}, \bibinfo {author} {\bibfnamefont
  {D.}~\bibnamefont {Teweldebrhan}}, \bibinfo {author} {\bibfnamefont
  {F.}~\bibnamefont {Miao}}, \ and\ \bibinfo {author} {\bibfnamefont {C.~N.}\
  \bibnamefont {Lau}},\ }\bibfield  {title} {\enquote {\bibinfo {title}
  {Superior thermal conductivity of single-layer graphene},}\ }\href {\doibase
  10.1021/nl0731872} {\bibfield  {journal} {\bibinfo  {journal} {Nano Lett.}\
  }\textbf {\bibinfo {volume} {8}},\ \bibinfo {pages} {902} (\bibinfo {year}
  {2008})}\BibitemShut {NoStop}%
\bibitem [{\citenamefont {Bonaccorso}\ \emph {et~al.}(2010)\citenamefont
  {Bonaccorso}, \citenamefont {Sun}, \citenamefont {Hasan},\ and\ \citenamefont
  {Ferrari}}]{Bonaccorso2010}%
  \BibitemOpen
  \bibfield  {author} {\bibinfo {author} {\bibfnamefont {F.}~\bibnamefont
  {Bonaccorso}}, \bibinfo {author} {\bibfnamefont {Z.}~\bibnamefont {Sun}},
  \bibinfo {author} {\bibfnamefont {T.}~\bibnamefont {Hasan}}, \ and\ \bibinfo
  {author} {\bibfnamefont {A.~C.}\ \bibnamefont {Ferrari}},\ }\bibfield
  {title} {\enquote {\bibinfo {title} {Graphene photonics and
  optoelectronics},}\ }\href {\doibase 10.1038/nphoton.2010.186} {\bibfield
  {journal} {\bibinfo  {journal} {Nat. Photon.}\ }\textbf {\bibinfo {volume}
  {4}},\ \bibinfo {pages} {611} (\bibinfo {year} {2010})}\BibitemShut {NoStop}%
\bibitem [{\citenamefont {Novoselov}\ \emph {et~al.}(2005)\citenamefont
  {Novoselov}, \citenamefont {Geim}, \citenamefont {Morozov}, \citenamefont
  {Jiang}, \citenamefont {Katsnelson}, \citenamefont {Grigorieva},
  \citenamefont {Dubonos},\ and\ \citenamefont {Firsov}}]{Novoselov2005}%
  \BibitemOpen
  \bibfield  {author} {\bibinfo {author} {\bibfnamefont {K.~S.}\ \bibnamefont
  {Novoselov}}, \bibinfo {author} {\bibfnamefont {A.~K.}\ \bibnamefont {Geim}},
  \bibinfo {author} {\bibfnamefont {S.~V.}\ \bibnamefont {Morozov}}, \bibinfo
  {author} {\bibfnamefont {D.}~\bibnamefont {Jiang}}, \bibinfo {author}
  {\bibfnamefont {M.~I.}\ \bibnamefont {Katsnelson}}, \bibinfo {author}
  {\bibfnamefont {I.~V.}\ \bibnamefont {Grigorieva}}, \bibinfo {author}
  {\bibfnamefont {S.~V.}\ \bibnamefont {Dubonos}}, \ and\ \bibinfo {author}
  {\bibfnamefont {A.~A.}\ \bibnamefont {Firsov}},\ }\bibfield  {title}
  {\enquote {\bibinfo {title} {Two-dimensional gas of massless {D}irac fermions
  in graphene},}\ }\href {\doibase 10.1038/nature04233} {\bibfield  {journal}
  {\bibinfo  {journal} {Nature}\ }\textbf {\bibinfo {volume} {438}},\ \bibinfo
  {pages} {197} (\bibinfo {year} {2005})}\BibitemShut {NoStop}%
\bibitem [{\citenamefont {Katsnelson}\ \emph {et~al.}(2006)\citenamefont
  {Katsnelson}, \citenamefont {Novoselov},\ and\ \citenamefont
  {Geim}}]{Katsnelson2006}%
  \BibitemOpen
  \bibfield  {author} {\bibinfo {author} {\bibfnamefont {M.~I.}\ \bibnamefont
  {Katsnelson}}, \bibinfo {author} {\bibfnamefont {K.~S.}\ \bibnamefont
  {Novoselov}}, \ and\ \bibinfo {author} {\bibfnamefont {A.~K.}\ \bibnamefont
  {Geim}},\ }\bibfield  {title} {\enquote {\bibinfo {title} {Chiral tunnelling
  and the {K}lein paradox in graphene},}\ }\href {\doibase 10.1038/nphys384}
  {\bibfield  {journal} {\bibinfo  {journal} {Nat. Phys.}\ }\textbf {\bibinfo
  {volume} {2}},\ \bibinfo {pages} {620} (\bibinfo {year} {2006})}\BibitemShut
  {NoStop}%
\bibitem [{\citenamefont {Novoselov}\ \emph {et~al.}(2007)\citenamefont
  {Novoselov}, \citenamefont {Jiang}, \citenamefont {Zhang}, \citenamefont
  {Morozov}, \citenamefont {Stormer}, \citenamefont {Zeitler}, \citenamefont
  {Maan}, \citenamefont {Boebinger}, \citenamefont {Kim},\ and\ \citenamefont
  {Geim}}]{Novoselov2007}%
  \BibitemOpen
  \bibfield  {author} {\bibinfo {author} {\bibfnamefont {K.~S.}\ \bibnamefont
  {Novoselov}}, \bibinfo {author} {\bibfnamefont {Z.}~\bibnamefont {Jiang}},
  \bibinfo {author} {\bibfnamefont {Y.}~\bibnamefont {Zhang}}, \bibinfo
  {author} {\bibfnamefont {S.~V.}\ \bibnamefont {Morozov}}, \bibinfo {author}
  {\bibfnamefont {H.~L.}\ \bibnamefont {Stormer}}, \bibinfo {author}
  {\bibfnamefont {U.}~\bibnamefont {Zeitler}}, \bibinfo {author} {\bibfnamefont
  {J.~C.}\ \bibnamefont {Maan}}, \bibinfo {author} {\bibfnamefont {G.~S.}\
  \bibnamefont {Boebinger}}, \bibinfo {author} {\bibfnamefont {P.}~\bibnamefont
  {Kim}}, \ and\ \bibinfo {author} {\bibfnamefont {A.~K.}\ \bibnamefont
  {Geim}},\ }\bibfield  {title} {\enquote {\bibinfo {title} {Room-temperature
  quantum {H}all effect in graphene},}\ }\href {\doibase
  10.1126/science.1137201} {\bibfield  {journal} {\bibinfo  {journal}
  {Science}\ }\textbf {\bibinfo {volume} {315}},\ \bibinfo {pages} {1379}
  (\bibinfo {year} {2007})}\BibitemShut {NoStop}%
\bibitem [{\citenamefont {Geim}\ and\ \citenamefont
  {MacDonald}(2007)}]{Geim2007}%
  \BibitemOpen
  \bibfield  {author} {\bibinfo {author} {\bibfnamefont {A.~K.}\ \bibnamefont
  {Geim}}\ and\ \bibinfo {author} {\bibfnamefont {A.~H.}\ \bibnamefont
  {MacDonald}},\ }\bibfield  {title} {\enquote {\bibinfo {title} {Graphene:
  Exploring carbon flatland},}\ }\href {\doibase 10.1063/1.2774096} {\bibfield
  {journal} {\bibinfo  {journal} {Phys. Today}\ }\textbf {\bibinfo {volume}
  {60}},\ \bibinfo {pages} {35} (\bibinfo {year} {2007})}\BibitemShut {NoStop}%
\bibitem [{\citenamefont {Das~Sarma}\ \emph {et~al.}(2011)\citenamefont
  {Das~Sarma}, \citenamefont {Adam}, \citenamefont {Hwang},\ and\ \citenamefont
  {Rossi}}]{DasSarma2011}%
  \BibitemOpen
  \bibfield  {author} {\bibinfo {author} {\bibfnamefont {S.}~\bibnamefont
  {Das~Sarma}}, \bibinfo {author} {\bibfnamefont {S.}~\bibnamefont {Adam}},
  \bibinfo {author} {\bibfnamefont {E.~H.}\ \bibnamefont {Hwang}}, \ and\
  \bibinfo {author} {\bibfnamefont {E.}~\bibnamefont {Rossi}},\ }\bibfield
  {title} {\enquote {\bibinfo {title} {Electronic transport in two-dimensional
  graphene},}\ }\href {\doibase 10.1103/RevModPhys.83.407} {\bibfield
  {journal} {\bibinfo  {journal} {Rev. Mod. Phys.}\ }\textbf {\bibinfo {volume}
  {83}},\ \bibinfo {pages} {407} (\bibinfo {year} {2011})}\BibitemShut
  {NoStop}%
\bibitem [{\citenamefont {Wallace}(1947)}]{Wallace1947}%
  \BibitemOpen
  \bibfield  {author} {\bibinfo {author} {\bibfnamefont {P.}~\bibnamefont
  {Wallace}},\ }\bibfield  {title} {\enquote {\bibinfo {title} {The band theory
  of graphite},}\ }\href {\doibase 10.1103/PhysRev.71.622} {\bibfield
  {journal} {\bibinfo  {journal} {Phys. Rev.}\ }\textbf {\bibinfo {volume}
  {71}},\ \bibinfo {pages} {622} (\bibinfo {year} {1947})}\BibitemShut
  {NoStop}%
\bibitem [{\citenamefont {Beenakker}(2008)}]{Beenakker2008a}%
  \BibitemOpen
  \bibfield  {author} {\bibinfo {author} {\bibfnamefont {C.~W.~J.}\
  \bibnamefont {Beenakker}},\ }\bibfield  {title} {\enquote {\bibinfo {title}
  {Colloquium: Andreev reflection and klein tunneling in graphene},}\ }\href
  {\doibase 10.1103/RevModPhys.80.1337} {\bibfield  {journal} {\bibinfo
  {journal} {Rev. Mod. Phys.}\ }\textbf {\bibinfo {volume} {80}},\ \bibinfo
  {eid} {1337} (\bibinfo {year} {2008})}\BibitemShut {NoStop}%
\bibitem [{\citenamefont {Goerbig}(2011)}]{Goerbig2011}%
  \BibitemOpen
  \bibfield  {author} {\bibinfo {author} {\bibfnamefont {M.~O.}\ \bibnamefont
  {Goerbig}},\ }\bibfield  {title} {\enquote {\bibinfo {title} {Electronic
  properties of graphene in a strong magnetic field},}\ }\href {\doibase
  10.1103/RevModPhys.83.1193} {\bibfield  {journal} {\bibinfo  {journal} {Rev.
  Mod. Phys.}\ }\textbf {\bibinfo {volume} {83}},\ \bibinfo {pages} {1193}
  (\bibinfo {year} {2011})}\BibitemShut {NoStop}%
\bibitem [{\citenamefont {Castro}\ \emph {et~al.}(2007)\citenamefont {Castro},
  \citenamefont {Novoselov}, \citenamefont {Morozov}, \citenamefont {Peres},
  \citenamefont {Lopes~dos Santos}, \citenamefont {Nilsson}, \citenamefont
  {Guinea}, \citenamefont {Geim},\ and\ \citenamefont
  {Castro~Neto}}]{castro_biased_2007}%
  \BibitemOpen
  \bibfield  {author} {\bibinfo {author} {\bibfnamefont {E.~V.}\ \bibnamefont
  {Castro}}, \bibinfo {author} {\bibfnamefont {K.~S.}\ \bibnamefont
  {Novoselov}}, \bibinfo {author} {\bibfnamefont {S.~V.}\ \bibnamefont
  {Morozov}}, \bibinfo {author} {\bibfnamefont {N.~M.~R.}\ \bibnamefont
  {Peres}}, \bibinfo {author} {\bibfnamefont {Lopes~J.~M.~B.}\ \bibnamefont
  {dos Santos}}, \bibinfo {author} {\bibfnamefont {J.}~\bibnamefont
  {Nilsson}}, \bibinfo {author} {\bibfnamefont {F.}~\bibnamefont {Guinea}},
  \bibinfo {author} {\bibfnamefont {A.~K.}\ \bibnamefont {Geim}}, \ and\
  \bibinfo {author} {\bibfnamefont {A.~H.}\ \bibnamefont {Castro~Neto}},\
  }\bibfield  {title} {\enquote {\bibinfo {title} {Biased {Bilayer Graphene}:
  {Semiconductor} with a {Gap Tunable} by the {Electric Field Effect}},}\
  }\href {\doibase 10.1103/PhysRevLett.99.216802} {\bibfield  {journal}
  {\bibinfo  {journal} {Phys. Rev. Lett.}\ }\textbf {\bibinfo {volume} {99}},\
  \bibinfo {pages} {216802} (\bibinfo {year} {2007})}\BibitemShut {NoStop}%
\bibitem [{\citenamefont {McCann}(2006)}]{McCann2006}%
  \BibitemOpen
  \bibfield  {author} {\bibinfo {author} {\bibfnamefont {E.}~\bibnamefont
  {McCann}},\ }\bibfield  {title} {\enquote {\bibinfo {title} {Asymmetry gap in
  the electronic band structure of bilayer graphene},}\ }\href {\doibase
  10.1103/PhysRevB.74.161403} {\bibfield  {journal} {\bibinfo  {journal} {Phys.
  Rev. B}\ }\textbf {\bibinfo {volume} {74}},\ \bibinfo {pages} {161403}
  (\bibinfo {year} {2006})}\BibitemShut {NoStop}%
\bibitem [{\citenamefont {Min}\ \emph {et~al.}(2007)\citenamefont {Min},
  \citenamefont {Sahu}, \citenamefont {Banerjee},\ and\ \citenamefont
  {MacDonald}}]{Min2007}%
  \BibitemOpen
  \bibfield  {author} {\bibinfo {author} {\bibfnamefont {H.}~\bibnamefont
  {Min}}, \bibinfo {author} {\bibfnamefont {B.}~\bibnamefont {Sahu}}, \bibinfo
  {author} {\bibfnamefont {S.~K.}\ \bibnamefont {Banerjee}}, \ and\ \bibinfo
  {author} {\bibfnamefont {A.~H.}\ \bibnamefont {MacDonald}},\ }\bibfield
  {title} {\enquote {\bibinfo {title} {Ab initio theory of gate induced gaps in
  graphene bilayers},}\ }\href {\doibase 10.1103/PhysRevB.75.155115} {\bibfield
   {journal} {\bibinfo  {journal} {Phys. Rev. B}\ }\textbf {\bibinfo {volume}
  {75}},\ \bibinfo {pages} {155115} (\bibinfo {year} {2007})}\BibitemShut
  {NoStop}%
\bibitem [{\citenamefont {Taychatanapat}\ and\ \citenamefont
  {Jarillo-Herrero}(2010)}]{Taychatanapat2010}%
  \BibitemOpen
  \bibfield  {author} {\bibinfo {author} {\bibfnamefont {T.}~\bibnamefont
  {Taychatanapat}}\ and\ \bibinfo {author} {\bibfnamefont {P.}~\bibnamefont
  {Jarillo-Herrero}},\ }\bibfield  {title} {\enquote {\bibinfo {title}
  {Electronic transport in dual-gated bilayer graphene at large displacement
  fields},}\ }\href {\doibase 10.1103/PhysRevLett.105.166601} {\bibfield
  {journal} {\bibinfo  {journal} {Phys. Rev. Lett.}\ }\textbf {\bibinfo
  {volume} {105}},\ \bibinfo {pages} {166601} (\bibinfo {year}
  {2010})}\BibitemShut {NoStop}%
\bibitem [{\citenamefont {Hasan}\ and\ \citenamefont {Kane}(2010)}]{Hasan2010}%
  \BibitemOpen
  \bibfield  {author} {\bibinfo {author} {\bibfnamefont {M.~Z.}\ \bibnamefont
  {Hasan}}\ and\ \bibinfo {author} {\bibfnamefont {C.~L.}\ \bibnamefont
  {Kane}},\ }\bibfield  {title} {\enquote {\bibinfo {title} {Colloquium:
  Topological insulators},}\ }\href {\doibase 10.1103/RevModPhys.82.3045}
  {\bibfield  {journal} {\bibinfo  {journal} {Rev. Mod. Phys.}\ }\textbf
  {\bibinfo {volume} {82}},\ \bibinfo {pages} {3045} (\bibinfo {year}
  {2010})}\BibitemShut {NoStop}%
\bibitem [{\citenamefont {Martin}\ \emph {et~al.}(2008)\citenamefont {Martin},
  \citenamefont {Blanter},\ and\ \citenamefont {Morpurgo}}]{Martin2008}%
  \BibitemOpen
  \bibfield  {author} {\bibinfo {author} {\bibfnamefont {I.}~\bibnamefont
  {Martin}}, \bibinfo {author} {\bibfnamefont {Y.~M.}\ \bibnamefont {Blanter}},
  \ and\ \bibinfo {author} {\bibfnamefont {A.~F.}\ \bibnamefont {Morpurgo}},\
  }\bibfield  {title} {\enquote {\bibinfo {title} {Topological confinement in
  bilayer graphene},}\ }\href {\doibase 10.1103/PhysRevLett.100.036804}
  {\bibfield  {journal} {\bibinfo  {journal} {Phys. Rev. Lett.}\ }\textbf
  {\bibinfo {volume} {100}},\ \bibinfo {pages} {036804} (\bibinfo {year}
  {2008})}\BibitemShut {NoStop}%
\bibitem [{\citenamefont {Rycerz}\ \emph {et~al.}(2007)\citenamefont {Rycerz},
  \citenamefont {Tworzyd{\l}o},\ and\ \citenamefont {Beenakker}}]{Rycerz2007}%
  \BibitemOpen
  \bibfield  {author} {\bibinfo {author} {\bibfnamefont {A.}~\bibnamefont
  {Rycerz}}, \bibinfo {author} {\bibfnamefont {J.}~\bibnamefont
  {Tworzyd{\l}o}}, \ and\ \bibinfo {author} {\bibfnamefont {C.~W.~J.}\
  \bibnamefont {Beenakker}},\ }\bibfield  {title} {\enquote {\bibinfo {title}
  {Valley filter and valley valve in graphene},}\ }\href {\doibase
  10.1038/nphys547} {\bibfield  {journal} {\bibinfo  {journal} {Nat. Phys.}\
  }\textbf {\bibinfo {volume} {3}},\ \bibinfo {pages} {172} (\bibinfo {year}
  {2007})}\BibitemShut {NoStop}%
\bibitem [{\citenamefont {Xiao}\ \emph {et~al.}(2007)\citenamefont {Xiao},
  \citenamefont {Yao},\ and\ \citenamefont
  {Niu}}]{xiao_valley-contrasting_2007}%
  \BibitemOpen
  \bibfield  {author} {\bibinfo {author} {\bibfnamefont {D.}~\bibnamefont
  {Xiao}}, \bibinfo {author} {\bibfnamefont {W.}~\bibnamefont {Yao}}, \ and\
  \bibinfo {author} {\bibfnamefont {Q.}~\bibnamefont {Niu}},\ }\bibfield
  {title} {\enquote {\bibinfo {title} {Valley-{Contrasting Physics} in
  {Graphene}: {Magnetic Moment} and {Topological Transport}},}\ }\href
  {\doibase 10.1103/PhysRevLett.99.236809} {\bibfield  {journal} {\bibinfo
  {journal} {Phys. Rev. Lett.}\ }\textbf {\bibinfo {volume} {99}},\ \bibinfo
  {pages} {236809} (\bibinfo {year} {2007})}\BibitemShut {NoStop}%
\bibitem [{\citenamefont {Sui}\ \emph {et~al.}(2015)\citenamefont {Sui},
  \citenamefont {Chen}, \citenamefont {Ma}, \citenamefont {Shan}, \citenamefont
  {Tian}, \citenamefont {Watanabe}, \citenamefont {Taniguchi}, \citenamefont
  {Jin}, \citenamefont {Yao}, \citenamefont {Xiao},\ and\ \citenamefont
  {Zhang}}]{Sui2015}%
  \BibitemOpen
  \bibfield  {author} {\bibinfo {author} {\bibfnamefont {M.}~\bibnamefont
  {Sui}}, \bibinfo {author} {\bibfnamefont {G.}~\bibnamefont {Chen}}, \bibinfo
  {author} {\bibfnamefont {L.}~\bibnamefont {Ma}}, \bibinfo {author}
  {\bibfnamefont {W.-Y.}\ \bibnamefont {Shan}}, \bibinfo {author}
  {\bibfnamefont {D.}~\bibnamefont {Tian}}, \bibinfo {author} {\bibfnamefont
  {K.}~\bibnamefont {Watanabe}}, \bibinfo {author} {\bibfnamefont
  {T.}~\bibnamefont {Taniguchi}}, \bibinfo {author} {\bibfnamefont
  {X.}~\bibnamefont {Jin}}, \bibinfo {author} {\bibfnamefont {W.}~\bibnamefont
  {Yao}}, \bibinfo {author} {\bibfnamefont {D.}~\bibnamefont {Xiao}}, \ and\
  \bibinfo {author} {\bibfnamefont {Y.}~\bibnamefont {Zhang}},\ }\bibfield
  {title} {\enquote {\bibinfo {title} {Gate-tunable topological valley
  transport in bilayer graphene},}\ }\href {\doibase 10.1038/nphys3485}
  {\bibfield  {journal} {\bibinfo  {journal} {Nat Phys}\ }\textbf {\bibinfo
  {volume} {11}},\ \bibinfo {pages} {1027} (\bibinfo {year}
  {2015})}\BibitemShut {NoStop}%
\bibitem [{\citenamefont {Xiao}\ \emph {et~al.}(2010)\citenamefont {Xiao},
  \citenamefont {Chang},\ and\ \citenamefont {Niu}}]{xiao_berry_2010}%
  \BibitemOpen
  \bibfield  {author} {\bibinfo {author} {\bibfnamefont {D.}~\bibnamefont
  {Xiao}}, \bibinfo {author} {\bibfnamefont {M.-C.}\ \bibnamefont {Chang}}, \
  and\ \bibinfo {author} {\bibfnamefont {Q.}~\bibnamefont {Niu}},\ }\bibfield
  {title} {\enquote {\bibinfo {title} {Berry phase effects on electronic
  properties},}\ }\href {\doibase 10.1103/RevModPhys.82.1959} {\bibfield
  {journal} {\bibinfo  {journal} {Rev. Mod. Phys.}\ }\textbf {\bibinfo {volume}
  {82}},\ \bibinfo {pages} {1959} (\bibinfo {year} {2010})}\BibitemShut
  {NoStop}%
\bibitem [{\citenamefont {Ando}(2013)}]{Ando2013}%
  \BibitemOpen
  \bibfield  {author} {\bibinfo {author} {\bibfnamefont {Y.}~\bibnamefont
  {Ando}},\ }\bibfield  {title} {\enquote {\bibinfo {title} {Topological
  insulator materials},}\ }\href {\doibase 10.7566/JPSJ.82.102001} {\bibfield
  {journal} {\bibinfo  {journal} {J. Phys. Soc. Jpn.}\ }\textbf {\bibinfo
  {volume} {82}},\ \bibinfo {pages} {102001} (\bibinfo {year}
  {2013})}\BibitemShut {NoStop}%
\bibitem [{\citenamefont {{Ren}}\ \emph {et~al.}(2015)\citenamefont {{Ren}},
  \citenamefont {{Qiao}},\ and\ \citenamefont {{Niu}}}]{Ren2015a}%
  \BibitemOpen
  \bibfield  {author} {\bibinfo {author} {\bibfnamefont {Y.}~\bibnamefont
  {{Ren}}}, \bibinfo {author} {\bibfnamefont {Z.}~\bibnamefont {{Qiao}}}, \
  and\ \bibinfo {author} {\bibfnamefont {Q.}~\bibnamefont {{Niu}}},\ }\bibfield
   {title} {\enquote {\bibinfo {title} {Topological phases in two-dimensional
  materials: A brief review},}\ }\href@noop {} {\bibfield  {journal} {\bibinfo
  {journal} {ArXiv e-prints}\ } (\bibinfo {year} {2015})},\ \Eprint
  {http://arxiv.org/abs/1509.09016} {arXiv:1509.09016} \BibitemShut {NoStop}%
\bibitem [{\citenamefont {Kechedzhi}\ \emph {et~al.}(2007)\citenamefont
  {Kechedzhi}, \citenamefont {Fal{\textquoteright}ko}, \citenamefont {McCann},\
  and\ \citenamefont {Altshuler}}]{tw_falko_altshuler_2007}%
  \BibitemOpen
  \bibfield  {author} {\bibinfo {author} {\bibfnamefont {K.}~\bibnamefont
  {Kechedzhi}}, \bibinfo {author} {\bibfnamefont {V.~I.}\ \bibnamefont
  {Fal{\textquoteright}ko}}, \bibinfo {author} {\bibfnamefont {E.}~\bibnamefont
  {McCann}}, \ and\ \bibinfo {author} {\bibfnamefont {B.~L.}\ \bibnamefont
  {Altshuler}},\ }\bibfield  {title} {\enquote {\bibinfo {title} {Influence of
  trigonal warping on interference effects in bilayer graphene},}\ }\href
  {\doibase 10.1103/PhysRevLett.98.176806} {\bibfield  {journal} {\bibinfo
  {journal} {Phys. Rev. Lett.}\ }\textbf {\bibinfo {volume} {98}},\ \bibinfo
  {pages} {176806} (\bibinfo {year} {2007})}\BibitemShut {NoStop}%
\bibitem [{\citenamefont {Cserti}\ \emph {et~al.}(2007)\citenamefont {Cserti},
  \citenamefont {Csord{\'{a}}s},\ and\ \citenamefont
  {D{\'{a}}vid}}]{tw_cserti_2007}%
  \BibitemOpen
  \bibfield  {author} {\bibinfo {author} {\bibfnamefont {J.}~\bibnamefont
  {Cserti}}, \bibinfo {author} {\bibfnamefont {A.}~\bibnamefont
  {Csord{\'{a}}s}}, \ and\ \bibinfo {author} {\bibfnamefont {G.}~\bibnamefont
  {D{\'{a}}vid}},\ }\bibfield  {title} {\enquote {\bibinfo {title} {Role of the
  trigonal warping on the minimal conductivity of bilayer graphene},}\ }\href
  {\doibase 10.1103/PhysRevLett.99.066802} {\bibfield  {journal} {\bibinfo
  {journal} {Phys. Rev. Lett.}\ }\textbf {\bibinfo {volume} {99}},\ \bibinfo
  {pages} {066802} (\bibinfo {year} {2007})}\BibitemShut {NoStop}%
\bibitem [{\citenamefont {Koshino}\ and\ \citenamefont
  {McCann}(2009)}]{tw_koshino_2009}%
  \BibitemOpen
  \bibfield  {author} {\bibinfo {author} {\bibfnamefont {M.}~\bibnamefont
  {Koshino}}\ and\ \bibinfo {author} {\bibfnamefont {E.}~\bibnamefont
  {McCann}},\ }\bibfield  {title} {\enquote {\bibinfo {title} {Trigonal warping
  and {B}erry{\textquoteright}s phase {N$\pi$} in {ABC}-stacked multilayer
  graphene},}\ }\href {\doibase 10.1103/PhysRevB.80.165409} {\bibfield
  {journal} {\bibinfo  {journal} {Phys. Rev. B.}\ }\textbf {\bibinfo {volume}
  {80}},\ \bibinfo {pages} {165409} (\bibinfo {year} {2009})}\BibitemShut
  {NoStop}%
\bibitem [{\citenamefont {Montambaux}\ \emph {et~al.}(2009)\citenamefont
  {Montambaux}, \citenamefont {Pi{\'{e}}chon}, \citenamefont {Fuchs},\ and\
  \citenamefont {Goerbig}}]{montambaux_merging_2009}%
  \BibitemOpen
  \bibfield  {author} {\bibinfo {author} {\bibfnamefont {G.}~\bibnamefont
  {Montambaux}}, \bibinfo {author} {\bibfnamefont {F.}~\bibnamefont
  {Pi{\'{e}}chon}}, \bibinfo {author} {\bibfnamefont {J.-N.}\ \bibnamefont
  {Fuchs}}, \ and\ \bibinfo {author} {\bibfnamefont {M.~O.}\ \bibnamefont
  {Goerbig}},\ }\bibfield  {title} {\enquote {\bibinfo {title} {Merging of
  {Dirac} points in a two-dimensional crystal},}\ }\href {\doibase
  10.1103/PhysRevB.80.153412} {\bibfield  {journal} {\bibinfo  {journal} {Phys.
  Rev. B}\ }\textbf {\bibinfo {volume} {80}},\ \bibinfo {pages} {153412}
  (\bibinfo {year} {2009})}\BibitemShut {NoStop}%
\bibitem [{\citenamefont {de~Gail}\ \emph {et~al.}(2012)\citenamefont
  {de~Gail}, \citenamefont {Goerbig},\ and\ \citenamefont
  {Montambaux}}]{de_gail_magnetic_2012}%
  \BibitemOpen
  \bibfield  {author} {\bibinfo {author} {\bibfnamefont {R.}~\bibnamefont
  {de~Gail}}, \bibinfo {author} {\bibfnamefont {M.~O.}\ \bibnamefont
  {Goerbig}}, \ and\ \bibinfo {author} {\bibfnamefont {G.}~\bibnamefont
  {Montambaux}},\ }\bibfield  {title} {\enquote {\bibinfo {title} {Magnetic
  spectrum of trigonally warped bilayer graphene: {Semiclassical} analysis,
  zero modes, and topological winding numbers},}\ }\href {\doibase
  10.1103/PhysRevB.86.045407} {\bibfield  {journal} {\bibinfo  {journal} {Phys.
  Rev. B}\ }\textbf {\bibinfo {volume} {86}},\ \bibinfo {pages} {045407}
  (\bibinfo {year} {2012})}\BibitemShut {NoStop}%
\bibitem [{\citenamefont {Fuchs}\ \emph {et~al.}(2010)\citenamefont {Fuchs},
  \citenamefont {Pi{\'{e}}chon}, \citenamefont {Goerbig},\ and\ \citenamefont
  {Montambaux}}]{fuchs_topological_2010}%
  \BibitemOpen
  \bibfield  {author} {\bibinfo {author} {\bibfnamefont {J.~N.}\ \bibnamefont
  {Fuchs}}, \bibinfo {author} {\bibfnamefont {F.}~\bibnamefont
  {Pi{\'{e}}chon}}, \bibinfo {author} {\bibfnamefont {M.~O.}\ \bibnamefont
  {Goerbig}}, \ and\ \bibinfo {author} {\bibfnamefont {G.}~\bibnamefont
  {Montambaux}},\ }\bibfield  {title} {\enquote {\bibinfo {title} {Topological
  {Berry} phase and semiclassical quantization of cyclotron orbits for two
  dimensional electrons in coupled band models},}\ }\href {\doibase
  10.1140/epjb/e2010-00259-2} {\bibfield  {journal} {\bibinfo  {journal} {Eur.
  Phys. J. B}\ }\textbf {\bibinfo {volume} {77}},\ \bibinfo {pages} {351}
  (\bibinfo {year} {2010})}\BibitemShut {NoStop}%
\bibitem [{\citenamefont {McMillan}(2002)}]{mcmillan_new_2002}%
  \BibitemOpen
  \bibfield  {author} {\bibinfo {author} {\bibfnamefont {P.~F.}\ \bibnamefont
  {McMillan}},\ }\bibfield  {title} {\enquote {\bibinfo {title} {New materials
  from high-pressure experiments},}\ }\href {\doibase 10.1038/nmat716}
  {\bibfield  {journal} {\bibinfo  {journal} {Nat. Mater.}\ }\textbf {\bibinfo
  {volume} {1}},\ \bibinfo {pages} {19} (\bibinfo {year} {2002})}\BibitemShut
  {NoStop}%
\bibitem [{\citenamefont {Novoselov}\ \emph {et~al.}(2006)\citenamefont
  {Novoselov}, \citenamefont {McCann}, \citenamefont {Morozov}, \citenamefont
  {Fal{\textquoteright}ko}, \citenamefont {Katsnelson}, \citenamefont
  {Zeitler}, \citenamefont {Jiang}, \citenamefont {Schedin},\ and\
  \citenamefont {Geim}}]{novoselov_unconventional_2006}%
  \BibitemOpen
  \bibfield  {author} {\bibinfo {author} {\bibfnamefont {K.~S.}\ \bibnamefont
  {Novoselov}}, \bibinfo {author} {\bibfnamefont {E.}~\bibnamefont {McCann}},
  \bibinfo {author} {\bibfnamefont {S.~V.}\ \bibnamefont {Morozov}}, \bibinfo
  {author} {\bibfnamefont {V.~I.}\ \bibnamefont {Fal{\textquoteright}ko}},
  \bibinfo {author} {\bibfnamefont {M.~I.}\ \bibnamefont {Katsnelson}},
  \bibinfo {author} {\bibfnamefont {U.}~\bibnamefont {Zeitler}}, \bibinfo
  {author} {\bibfnamefont {D.}~\bibnamefont {Jiang}}, \bibinfo {author}
  {\bibfnamefont {F.}~\bibnamefont {Schedin}}, \ and\ \bibinfo {author}
  {\bibfnamefont {A.~K.}\ \bibnamefont {Geim}},\ }\bibfield  {title} {\enquote
  {\bibinfo {title} {Unconventional quantum {Hall} effect and
  {Berry}{\textquoteright}s phase of 2$\pi$ in bilayer graphene},}\ }\href
  {\doibase 10.1038/nphys245} {\bibfield  {journal} {\bibinfo  {journal} {Nat
  Phys}\ }\textbf {\bibinfo {volume} {2}},\ \bibinfo {pages} {177} (\bibinfo
  {year} {2006})}\BibitemShut {NoStop}%
\bibitem [{\citenamefont {Kresse}\ and\ \citenamefont
  {Hafner}(1993)}]{kresse_ab_1993}%
  \BibitemOpen
  \bibfield  {author} {\bibinfo {author} {\bibfnamefont {G.}~\bibnamefont
  {Kresse}}\ and\ \bibinfo {author} {\bibfnamefont {J.}~\bibnamefont
  {Hafner}},\ }\bibfield  {title} {\enquote {\bibinfo {title} {Ab initio
  molecular dynamics for liquid metals},}\ }\href {\doibase
  10.1103/PhysRevB.47.558} {\bibfield  {journal} {\bibinfo  {journal} {Phys.
  Rev. B}\ }\textbf {\bibinfo {volume} {47}},\ \bibinfo {pages} {558} (\bibinfo
  {year} {1993})}\BibitemShut {NoStop}%
\bibitem [{\citenamefont {Kresse}\ and\ \citenamefont
  {Hafner}(1994)}]{kresse_ab_1994}%
  \BibitemOpen
  \bibfield  {author} {\bibinfo {author} {\bibfnamefont {G.}~\bibnamefont
  {Kresse}}\ and\ \bibinfo {author} {\bibfnamefont {J.}~\bibnamefont
  {Hafner}},\ }\bibfield  {title} {\enquote {\bibinfo {title} {Ab initio
  molecular-dynamics simulation of the liquid-metal--amorphous-semiconductor
  transition in germanium},}\ }\href {\doibase 10.1103/PhysRevB.49.14251}
  {\bibfield  {journal} {\bibinfo  {journal} {Phys. Rev. B}\ }\textbf {\bibinfo
  {volume} {49}},\ \bibinfo {pages} {14251} (\bibinfo {year}
  {1994})}\BibitemShut {NoStop}%
\bibitem [{\citenamefont {Kresse}\ and\ \citenamefont
  {Furthm{\"{u}}ller}(1996{\natexlab{a}})}]{kresse_efficiency_1996}%
  \BibitemOpen
  \bibfield  {author} {\bibinfo {author} {\bibfnamefont {G.}~\bibnamefont
  {Kresse}}\ and\ \bibinfo {author} {\bibfnamefont {J.}~\bibnamefont
  {Furthm{\"{u}}ller}},\ }\bibfield  {title} {\enquote {\bibinfo {title}
  {Efficiency of ab-initio total energy calculations for metals and
  semiconductors using a plane-wave basis set},}\ }\href {\doibase
  10.1016/0927-0256(96)00008-0} {\bibfield  {journal} {\bibinfo  {journal}
  {Comput. Mater. Sci.}\ }\textbf {\bibinfo {volume} {6}},\ \bibinfo {pages}
  {15} (\bibinfo {year} {1996}{\natexlab{a}})}\BibitemShut {NoStop}%
\bibitem [{\citenamefont {Kresse}\ and\ \citenamefont
  {Furthm{\"{u}}ller}(1996{\natexlab{b}})}]{kresse_efficient_1996}%
  \BibitemOpen
  \bibfield  {author} {\bibinfo {author} {\bibfnamefont {G.}~\bibnamefont
  {Kresse}}\ and\ \bibinfo {author} {\bibfnamefont {J.}~\bibnamefont
  {Furthm{\"{u}}ller}},\ }\bibfield  {title} {\enquote {\bibinfo {title}
  {Efficient iterative schemes for ab initio total-energy calculations using a
  plane-wave basis set},}\ }\href {\doibase 10.1103/PhysRevB.54.11169}
  {\bibfield  {journal} {\bibinfo  {journal} {Phys. Rev. B}\ }\textbf {\bibinfo
  {volume} {54}},\ \bibinfo {pages} {11169} (\bibinfo {year}
  {1996}{\natexlab{b}})}\BibitemShut {NoStop}%
\bibitem [{\citenamefont {Bl{\"{o}}chl}(1994)}]{blochl_projector_1994}%
  \BibitemOpen
  \bibfield  {author} {\bibinfo {author} {\bibfnamefont {P.~E.}\ \bibnamefont
  {Bl{\"{o}}chl}},\ }\bibfield  {title} {\enquote {\bibinfo {title} {Projector
  augmented-wave method},}\ }\href {\doibase 10.1103/PhysRevB.50.17953}
  {\bibfield  {journal} {\bibinfo  {journal} {Phys. Rev. B}\ }\textbf {\bibinfo
  {volume} {50}},\ \bibinfo {pages} {17953} (\bibinfo {year}
  {1994})}\BibitemShut {NoStop}%
\bibitem [{\citenamefont {Kresse}\ and\ \citenamefont
  {Joubert}(1999)}]{kresse_ultrasoft_1999}%
  \BibitemOpen
  \bibfield  {author} {\bibinfo {author} {\bibfnamefont {G.}~\bibnamefont
  {Kresse}}\ and\ \bibinfo {author} {\bibfnamefont {D.}~\bibnamefont
  {Joubert}},\ }\bibfield  {title} {\enquote {\bibinfo {title} {From ultrasoft
  pseudopotentials to the projector augmented-wave method},}\ }\href {\doibase
  10.1103/PhysRevB.59.1758} {\bibfield  {journal} {\bibinfo  {journal} {Phys.
  Rev. B}\ }\textbf {\bibinfo {volume} {59}},\ \bibinfo {pages} {1758}
  (\bibinfo {year} {1999})}\BibitemShut {NoStop}%
\bibitem [{\citenamefont {Perdew}\ \emph {et~al.}(1996)\citenamefont {Perdew},
  \citenamefont {Burke},\ and\ \citenamefont
  {Ernzerhof}}]{perdew_generalized_1996}%
  \BibitemOpen
  \bibfield  {author} {\bibinfo {author} {\bibfnamefont {J.~P.}\ \bibnamefont
  {Perdew}}, \bibinfo {author} {\bibfnamefont {K.}~\bibnamefont {Burke}}, \
  and\ \bibinfo {author} {\bibfnamefont {M.}~\bibnamefont {Ernzerhof}},\
  }\bibfield  {title} {\enquote {\bibinfo {title} {Generalized {Gradient
  Approximation Made Simple}},}\ }\href {\doibase 10.1103/PhysRevLett.77.3865}
  {\bibfield  {journal} {\bibinfo  {journal} {Phys. Rev. Lett.}\ }\textbf
  {\bibinfo {volume} {77}},\ \bibinfo {pages} {3865} (\bibinfo {year}
  {1996})}\BibitemShut {NoStop}%
\bibitem [{\citenamefont {Perdew}\ \emph {et~al.}(1997)\citenamefont {Perdew},
  \citenamefont {Burke},\ and\ \citenamefont
  {Ernzerhof}}]{perdew_generalized_1997}%
  \BibitemOpen
  \bibfield  {author} {\bibinfo {author} {\bibfnamefont {J.~P.}\ \bibnamefont
  {Perdew}}, \bibinfo {author} {\bibfnamefont {K.}~\bibnamefont {Burke}}, \
  and\ \bibinfo {author} {\bibfnamefont {M.}~\bibnamefont {Ernzerhof}},\
  }\bibfield  {title} {\enquote {\bibinfo {title} {Generalized {Gradient
  Approximation Made Simple} [{Phys}. {Rev}. {Lett}. 77, 3865 (1996)]},}\
  }\href {\doibase 10.1103/PhysRevLett.78.1396} {\bibfield  {journal} {\bibinfo
   {journal} {Phys. Rev. Lett.}\ }\textbf {\bibinfo {volume} {78}},\ \bibinfo
  {pages} {1396} (\bibinfo {year} {1997})}\BibitemShut {NoStop}%
\bibitem [{Note1()}]{Note1}%
  \BibitemOpen
  \bibinfo {note} {It should be emphasize that trigonal warping effects are
  always present, even a zero pressure. However, in such a case they occur at
  much lower energy, since the $t_3$ and $t_4$ hopping parameters (see next
  section) are rather small. In the first approximation, we take this limit as
  if $t_3=t_4=0$.}\BibitemShut {Stop}%
\bibitem [{\citenamefont {McCann}\ \emph {et~al.}(2007)\citenamefont {McCann},
  \citenamefont {Abergel},\ and\ \citenamefont
  {Fal{\textquoteright}ko}}]{mccann_electrons_2007}%
  \BibitemOpen
  \bibfield  {author} {\bibinfo {author} {\bibfnamefont {E.}~\bibnamefont
  {McCann}}, \bibinfo {author} {\bibfnamefont {D.~S.~L.}\ \bibnamefont
  {Abergel}}, \ and\ \bibinfo {author} {\bibfnamefont {V.~I.}\ \bibnamefont
  {Fal{\textquoteright}ko}},\ }\bibfield  {title} {\enquote {\bibinfo {title}
  {Electrons in bilayer graphene},}\ }\href {\doibase
  10.1016/j.ssc.2007.03.054} {\bibfield  {journal} {\bibinfo  {journal} {Solid
  State Commun.}\ }\textbf {\bibinfo {volume} {143}},\ \bibinfo {pages} {110}
  (\bibinfo {year} {2007})}\BibitemShut {NoStop}%
\bibitem [{\citenamefont {Nilsson}\ \emph {et~al.}(2008)\citenamefont
  {Nilsson}, \citenamefont {Castro~Neto}, \citenamefont {Guinea},\ and\
  \citenamefont {Peres}}]{nilsson_electronic_2008}%
  \BibitemOpen
  \bibfield  {author} {\bibinfo {author} {\bibfnamefont {J.}~\bibnamefont
  {Nilsson}}, \bibinfo {author} {\bibfnamefont {A.~H.}\ \bibnamefont
  {Castro~Neto}}, \bibinfo {author} {\bibfnamefont {F.}~\bibnamefont {Guinea}},
  \ and\ \bibinfo {author} {\bibfnamefont {N.~M.~R.}\ \bibnamefont {Peres}},\
  }\bibfield  {title} {\enquote {\bibinfo {title} {Electronic properties of
  bilayer and multilayer graphene},}\ }\href {\doibase
  10.1103/PhysRevB.78.045405} {\bibfield  {journal} {\bibinfo  {journal} {Phys.
  Rev. B}\ }\textbf {\bibinfo {volume} {78}},\ \bibinfo {pages} {045405}
  (\bibinfo {year} {2008})}\BibitemShut {NoStop}%
\bibitem [{\citenamefont {Konschuh}\ \emph {et~al.}(2012)\citenamefont
  {Konschuh}, \citenamefont {Gmitra}, \citenamefont {Kochan},\ and\
  \citenamefont {Fabian}}]{konschuh_theory_2012}%
  \BibitemOpen
  \bibfield  {author} {\bibinfo {author} {\bibfnamefont {S.}~\bibnamefont
  {Konschuh}}, \bibinfo {author} {\bibfnamefont {M.}~\bibnamefont {Gmitra}},
  \bibinfo {author} {\bibfnamefont {D.}~\bibnamefont {Kochan}}, \ and\ \bibinfo
  {author} {\bibfnamefont {J.}~\bibnamefont {Fabian}},\ }\bibfield  {title}
  {\enquote {\bibinfo {title} {Theory of spin-orbit coupling in bilayer
  graphene},}\ }\href {\doibase 10.1103/PhysRevB.85.115423} {\bibfield
  {journal} {\bibinfo  {journal} {Phys. Rev. B}\ }\textbf {\bibinfo {volume}
  {85}},\ \bibinfo {pages} {115423} (\bibinfo {year} {2012})}\BibitemShut
  {NoStop}%
\bibitem [{\citenamefont {Zou}\ \emph {et~al.}(2011)\citenamefont {Zou},
  \citenamefont {Hong},\ and\ \citenamefont {Zhu}}]{Zou11}%
  \BibitemOpen
  \bibfield  {author} {\bibinfo {author} {\bibfnamefont {K.}~\bibnamefont
  {Zou}}, \bibinfo {author} {\bibfnamefont {X.}~\bibnamefont {Hong}}, \ and\
  \bibinfo {author} {\bibfnamefont {J.}~\bibnamefont {Zhu}},\ }\bibfield
  {title} {\enquote {\bibinfo {title} {Effective mass of electrons and holes in
  bilayer graphene: Electron-hole asymmetry and electron-electron
  interaction},}\ }\href {\doibase 10.1103/PhysRevB.84.085408} {\bibfield
  {journal} {\bibinfo  {journal} {Phys. Rev. B}\ }\textbf {\bibinfo {volume}
  {84}},\ \bibinfo {pages} {085408} (\bibinfo {year} {2011})}\BibitemShut
  {NoStop}%
\bibitem [{\citenamefont {Zhang}\ \emph {et~al.}(2008)\citenamefont {Zhang},
  \citenamefont {Li}, \citenamefont {Basov}, \citenamefont {Fogler},
  \citenamefont {Hao},\ and\ \citenamefont
  {Martin}}]{zhang_determination_2008}%
  \BibitemOpen
  \bibfield  {author} {\bibinfo {author} {\bibfnamefont {L.~M.}\ \bibnamefont
  {Zhang}}, \bibinfo {author} {\bibfnamefont {Z.~Q.}\ \bibnamefont {Li}},
  \bibinfo {author} {\bibfnamefont {D.~N.}\ \bibnamefont {Basov}}, \bibinfo
  {author} {\bibfnamefont {M.~M.}\ \bibnamefont {Fogler}}, \bibinfo {author}
  {\bibfnamefont {Z.}~\bibnamefont {Hao}}, \ and\ \bibinfo {author}
  {\bibfnamefont {M.~C.}\ \bibnamefont {Martin}},\ }\bibfield  {title}
  {\enquote {\bibinfo {title} {Determination of the electronic structure of
  bilayer graphene from infrared spectroscopy},}\ }\href {\doibase
  10.1103/PhysRevB.78.235408} {\bibfield  {journal} {\bibinfo  {journal} {Phys.
  Rev. B}\ }\textbf {\bibinfo {volume} {78}},\ \bibinfo {pages} {235408}
  (\bibinfo {year} {2008})}\BibitemShut {NoStop}%
\bibitem [{\citenamefont {Malard}\ \emph {et~al.}(2007)\citenamefont {Malard},
  \citenamefont {Nilsson}, \citenamefont {Elias}, \citenamefont {Brant},
  \citenamefont {Plentz}, \citenamefont {Alves}, \citenamefont {Castro~Neto},\
  and\ \citenamefont {Pimenta}}]{Malard07}%
  \BibitemOpen
  \bibfield  {author} {\bibinfo {author} {\bibfnamefont {L.~M.}\ \bibnamefont
  {Malard}}, \bibinfo {author} {\bibfnamefont {J.}~\bibnamefont {Nilsson}},
  \bibinfo {author} {\bibfnamefont {D.~C.}\ \bibnamefont {Elias}}, \bibinfo
  {author} {\bibfnamefont {J.~C.}\ \bibnamefont {Brant}}, \bibinfo {author}
  {\bibfnamefont {F.}~\bibnamefont {Plentz}}, \bibinfo {author} {\bibfnamefont
  {E.~S.}\ \bibnamefont {Alves}}, \bibinfo {author} {\bibfnamefont {A.~H.}\
  \bibnamefont {Castro~Neto}}, \ and\ \bibinfo {author} {\bibfnamefont {M.~A.}\
  \bibnamefont {Pimenta}},\ }\bibfield  {title} {\enquote {\bibinfo {title}
  {Probing the electronic structure of bilayer graphene by {R}aman
  scattering},}\ }\href {\doibase 10.1103/PhysRevB.76.201401} {\bibfield
  {journal} {\bibinfo  {journal} {Phys. Rev. B}\ }\textbf {\bibinfo {volume}
  {76}},\ \bibinfo {pages} {201401} (\bibinfo {year} {2007})}\BibitemShut
  {NoStop}%
\bibitem [{Note2()}]{Note2}%
  \BibitemOpen
  \bibinfo {note} {We have compared the result of calculating the Berry
  curvature from the $4\times 4$ model and the $2\times 2$ approximation and
  the difference is negligible.}\BibitemShut {Stop}%
\bibitem [{\citenamefont {Yin}\ \emph {et~al.}(2015)\citenamefont {Yin},
  \citenamefont {Zhang}, \citenamefont {Qiao}, \citenamefont {Li},\ and\
  \citenamefont {He}}]{yin_experimental_2015}%
  \BibitemOpen
  \bibfield  {author} {\bibinfo {author} {\bibfnamefont {L.-J.}\ \bibnamefont
  {Yin}}, \bibinfo {author} {\bibfnamefont {Y.}~\bibnamefont {Zhang}}, \bibinfo
  {author} {\bibfnamefont {J.-B.}\ \bibnamefont {Qiao}}, \bibinfo {author}
  {\bibfnamefont {S.-Y.}\ \bibnamefont {Li}}, \ and\ \bibinfo {author}
  {\bibfnamefont {L.}~\bibnamefont {He}},\ }\bibfield  {title} {\enquote
  {\bibinfo {title} {Experimental {Observation} of {Surface States} and {Landau
  Levels Bending} in {Bilayer Graphene}},}\ }\href
  {http://arxiv.org/abs/1510.06109} {\bibfield  {journal} {\bibinfo  {journal}
  {arXiv:1510.06109 [cond-mat]}\ } (\bibinfo {year} {2015})},\ \bibinfo {note}
  {arXiv: 1510.06109}\BibitemShut {NoStop}%
\bibitem [{\citenamefont {Kawarabayashi}\ \emph {et~al.}(2013)\citenamefont
  {Kawarabayashi}, \citenamefont {Hasugai},\ and\ \citenamefont
  {Aoki}}]{kawarabayashi_stability_2013}%
  \BibitemOpen
  \bibfield  {author} {\bibinfo {author} {\bibfnamefont {T.}~\bibnamefont
  {Kawarabayashi}}, \bibinfo {author} {\bibfnamefont {Y.}~\bibnamefont
  {Hasugai}}, \ and\ \bibinfo {author} {\bibfnamefont {H.}~\bibnamefont
  {Aoki}},\ }\bibfield  {title} {\enquote {\bibinfo {title} {Stability of
  zero-mode {Landau} levels in bilayer graphene against disorder in the
  presence of the trigonal warping},}\ }\href {\doibase
  10.1088/1742-6596/456/1/012020} {\bibfield  {journal} {\bibinfo  {journal}
  {J. Phys.: Conf. Ser.}\ }\textbf {\bibinfo {volume} {456}},\ \bibinfo {pages}
  {012020} (\bibinfo {year} {2013})}\BibitemShut {NoStop}%
\bibitem [{\citenamefont {Yuan}\ \emph {et~al.}(2011)\citenamefont {Yuan},
  \citenamefont {Rold{\'{a}}n},\ and\ \citenamefont
  {Katsnelson}}]{yuan_landau_2011}%
  \BibitemOpen
  \bibfield  {author} {\bibinfo {author} {\bibfnamefont {S.}~\bibnamefont
  {Yuan}}, \bibinfo {author} {\bibfnamefont {R.}~\bibnamefont {Rold{\'{a}}n}},
  \ and\ \bibinfo {author} {\bibfnamefont {M.~I.}\ \bibnamefont {Katsnelson}},\
  }\bibfield  {title} {\enquote {\bibinfo {title} {Landau level spectrum of
  {ABA}- and {ABC}-stacked trilayer graphene},}\ }\href {\doibase
  10.1103/PhysRevB.84.125455} {\bibfield  {journal} {\bibinfo  {journal} {Phys.
  Rev. B}\ }\textbf {\bibinfo {volume} {84}},\ \bibinfo {pages} {125455}
  (\bibinfo {year} {2011})}\BibitemShut {NoStop}%
\bibitem [{\citenamefont {Onsager}(1952)}]{onsager_interpretation_1952}%
  \BibitemOpen
  \bibfield  {author} {\bibinfo {author} {\bibfnamefont {L.}~\bibnamefont
  {Onsager}},\ }\bibfield  {title} {\enquote {\bibinfo {title} {Interpretation
  of the de {Haas}-van {Alphen Effect}},}\ }\href {\doibase
  10.1080/14786440908521019} {\bibfield  {journal} {\bibinfo  {journal} {Phil.
  Mag.}\ }\textbf {\bibinfo {volume} {43}},\ \bibinfo {pages} {1006} (\bibinfo
  {year} {1952})}\BibitemShut {NoStop}%
\end{thebibliography}
\end{document}